\documentclass{article}
\usepackage[utf8]{inputenc}
\usepackage{graphicx}
\usepackage{color}
\usepackage{multirow}
\usepackage{latexsym}
\usepackage{amsmath,array}
\usepackage{tabularx}
\usepackage{soul}
\usepackage{tabu}

\title{Completing the Quantum Reconstruction Program via the Relativity Principle}

\author{W.M. Stuckey\thanks{Department of Physics, Elizabethtown College, Elizabethtown, PA 17022, USA} \, Michael Silberstein\thanks{Department of Philosophy, Elizabethtown College, Elizabethtown, PA 17022, USA} \, \thanks{Department of Philosophy, University of Maryland, College Park, MD 20742, USA}\, and Timothy McDevitt\thanks{Department of Mathematical Science, Elizabethtown College, Elizabethtown, PA 17022, USA}}

\date{}

\begin{document}

\maketitle
\sloppy
\begin{abstract}
  \noindent We explain how the disparate kinematics of quantum mechanics (finite-dimensional Hilbert space of QM) and special relativity (Minkowski spacetime from the Lorentz transformations of SR) can both be based on one principle (relativity principle). This is made possible by the axiomatic reconstruction of QM via information-theoretic principles, which has successfully recast QM as a \textit{principle theory} a la SR. That is, in the quantum reconstruction program (QRP) and SR, the formalisms (Hilbert space and Lorentz transformations, respectively) are derived from empirically discovered facts (Information Invariance \& Continuity and light postulate, respectively), so QM and SR are ``principle theories'' as defined by Einstein. While SR has a compelling fundamental principle to justify its empirically discovered fact (relativity principle), QRP has not produced a compelling fundamental principle or causal mechanism to account for its empirically discovered fact. To unify these disparate kinematics, we show how the relativity principle (``no preferred reference frame'' NPRF) can also be used to justify Information Invariance \& Continuity. We do this by showing that when QRP's operational notion of measurement is spatialized, Information Invariance \& Continuity entails the empirically discovered fact that everyone measures the same value for Planck's constant $h$, regardless of their relative spatial orientations or locations (Planck postulate). Since Poincar\'e transformations relate inertial reference frames via spatial rotations and translations as well as boosts, and Planck's constant $h$ and the speed of light $c$ are constants of Nature per Planck's radiation law and Maxwell's equations (respectively), the relativity principle justifies the Planck postulate just like it justifies the light postulate. Essentially, NPRF + $c$ is an adynamical global constraint over the spacetime configuration of worldtubes for bodily objects while NPRF + $h$ is an adynamical global constraint over the distribution of quanta among those bodily objects.
\end{abstract}


\section{Introduction}\label{SectionIntro}

Clauser won the 2022 Nobel Prize in Physics for his work in quantum physics and said \cite{GreenEntanglement}:
\begin{quote}
I was, again, very saddened that I had not overthrown quantum mechanics because I had, and to this day still have, great difficulty in understanding it.
\end{quote}
Gell-Mann also won the Nobel Prize in Physics for his work in quantum physics and said \cite[p. 144]{gell-mann}:
\begin{quote}
    We all know how to use [quantum mechanics] and how to apply it to problems; and so we have learned to live with the fact that nobody can understand it.
\end{quote}
Feynman is yet another Nobel Prize winner for his work in quantum physics who said \cite{feynmanQM2}:
\begin{quote}
    I think I can safely say that nobody understands quantum mechanics.
\end{quote}
Despite the fact that ``There is no ambiguity, no confusion, and spectacular success'' when we use quantum mechanics, ``we lack ... any consensus about what one is actually \textit{talking about} as one uses quantum mechanics'' \cite{mermin2019}. This obtains despite decades of work trying to interpret quantum mechanics (QM). 

While the foundations of physics community has produced many interpretations of QM, none has found consensus support. \cite{fuchsQMasQI} writes:
\begin{quote}
    What is the cause of this year-after-year sacrifice to the ``great mystery?'' Whatever it is, it cannot be for want of a self-ordained solution: Go to any meeting, and it is like being in a holy city in great tumult. You will find all the religions with all their priests pitted in holy war ... . They all declare to see the light, the ultimate light. Each tells us that if we will accept their solution as our savior, then we too will see the light.
\end{quote}
As \cite{maudlinQNLandSR} points out, every interpretation has its flaw that the other ``religions'' can attack:
\begin{quote}
    They may correctly note that according to every one of their rival theories, God was malicious, and having thus eliminated every other possibility, claim their own theory the victor. The problem is that \textit{every} partisan can argue in this way since \textit{every} theory posits some funny business on the part of the Deity.
\end{quote}

\clearpage

The problem as articulated by \cite{vancamp2011} is that, ``Constructive interpretations are attempted, but they are not unequivocally constructive in any traditional sense.'' \cite{vancamp2011} concludes:
\begin{quote}
    The interpretive work that must be done is less in coming up with a constructive theory and thereby explaining puzzling quantum phenomena, but more in explaining why the interpretation counts as explanatory at all given that it must sacrifice some key aspect of the traditional understanding of causal-mechanical explanation.
\end{quote}
In short, despite decades of effort within the foundations community no consensus constructive account of QM has been produced. Nor for that matter has any consensus principle or structural account of QM been produced.

According to Einstein, a constructive theory is based on dynamical laws and/or mechanistic causal processes (causal mechanisms) while a principle theory is based on an empirically discovered fact \cite{einstein1919}. He used the kinetic theory of gases as an example of a constructive theory and thermodynamics as an example of a principle theory where the empirically discovered fact at its foundation is ``perpetual motion machines are impossible.'' Special relativity (SR) is also a principle theory based on the empirically discovered fact that everyone measures the same value for the speed of light $c$, regardless of their uniform relative motions (light postulate). Another way to view the distinction between constructive and principle theories was proposed by \cite{maltranaEtAl2022}:
\begin{quote}
    Those theories that allow us to \textit{trace} the \textit{causal mechanisms} that \textit{explain mechanistically} the occurrence of a certain phenomenon we call ``mechanistic theories.'' And those theories that lack agents whose actions are causally responsible for phenomena, but that instead provide general constraints or structural elements that lead to unificationist explanations we call ``structural theories.''
\end{quote}
Obviously, their mechanistic theory corresponds to Einstein's constructive theory and their structural theory corresponds to Einstein's principle theory. According to the conventional definition of constructive or mechanistic explanation, causal mechanisms provide the explanans. By extension, one would then expect principle explanation to be the same as structural explanation, but in Section \ref{SectionExplanation} we propose an important distinction. 

Briefly, in structural explanation per \cite{Felline2018Synthese} the formalism of the principle theory provides the explanans while in principle explanation per \cite{silberstein2021} a compelling fundamental principle that justifies the empirically discovered fact at the basis of a principle theory provides the most fundamental explanans. For example, \cite{Felline2018Synthese} and \cite{Janssen2009} would say the explanans for structural explanation in the principle theory of SR is the Minkowski spacetime (M4) of its formalism (Lorentz transformations) and the invariant causal structure of M4 is so widely accepted as an explanans that it is commonly seen in presentations for a general audience, e.g., \cite{PBScausalM4}. Notice that this does not require any reason for the existence of the empirically discovered fact, whence the formalism of the corresponding principle theory. 

In SR, the justification of its empirically discovered fact (observer-independence of $c$) is the relativity principle -- the laws of physics (including their constants of Nature) are the same in all inertial reference frames -- because the value $c$ in the light postulate is part of Maxwell's equations. We label this NPRF + $c$ for short, where NPRF stands for ``no preferred reference frame.'' So, \cite{silberstein2021} would say the most fundamental explanans for principle explanation in SR is the relativity principle. This principle explanation is so widely accepted that it is presented in popular introductory physics textbooks, e.g., \cite{serway} and \cite{knight}, without controversy or qualification.  

Of course, as \cite{norton2004} and \cite{DarrigolBook2022} point out, the relativity principle was important for Newtonian mechanics as well. Indeed, as explained by \cite{Goyal2020}, Newton's laws of motion can be derived from conservation and the relativity principle. The difference is that Newtonian mechanics and SR are based on two different empirically discovered facts, i.e., the speed of light is infinity in Newtonian mechanics while it is $c$ in SR. In Newtonian mechanics, that empirically discovered fact is unjustified while for SR it follows necessarily from Maxwell's equations. \cite{norton2004} writes:
\begin{quote}
Until this electrodynamics emerged, special relativity could not arise; once it had emerged, special relativity could not be stopped. 
\end{quote}  
Once you have SR, Newtonian mechanics is understood as its $c \rightarrow \infty$ limit. Per \cite{silberstein2021}, by justifying the light postulate the relativity principle is ultimately responsible for reconciling a conflict between Galilean-invariant Newtonian mechanics and Lorentz-invariant Maxwell's equations, thereby unifying mechanics and electrodynamics by ``a subsumption of both theories under a higher principle'' \cite{Giovanelli2023}.

Our goal here is to bring this distinction to bear on the information-theoretic reconstruction of QM (the quantum reconstruction program, QRP), which has succeeded in rendering QM a principle theory without producing a consensus understanding of QM. We argue that the reason (the otherwise successful) QRP has failed to produce a consensus understanding of QM is that structural explanation per its formalism of information-theoretic principles fails to provide either of the two characteristics of a successful ``principle theory approach'' per \cite{vancamp2011}, i.e., it has failed to provide ``clear empirical understanding'' or increase unification. So, structural explanation fails for the reconstruction of QM even though it is successful and reasonable for SR. Therefore, we propose a completion of QRP via principle explanation per \cite{silberstein2021} that does result in ``clear empirical understanding'' and broad-based unification. This completion may be thought of as an ``interpretation of QRP'' via the spatialization of QRP's purely operational notion of measurement per \cite{GoyalPhenomQBism2024}. And perhaps surprisingly, that unification is between QM and SR, two theories that are widely believed to be in tension or outright conflict due to quantum entanglement \cite{albertGalchen2009,mamone}. 

\noindent For example, \cite{bellSR1986} complained:
\begin{quote}
    For me this is the real problem with quantum theory: the apparently essential conflict between any sharp formulation and fundamental relativity. 
\end{quote}
Likewise, \cite{PR1994} write:
\begin{quote}
    Quantum mechanics, which does not allow us to transmit signals faster than light, preserves relativistic causality. But quantum mechanics does not always allow us to consider distant systems as separate, as Einstein assumed. The failure of Einstein separability violates, not the letter, but the spirit of special relativity, and left many physicists (including Bell) deeply unsettled.
\end{quote}
And, \cite[p. 23]{maudlinQNLandSR} writes:
\begin{quote}
    We cannot simply accept the pronouncements of our best theories, no matter how strange, if those pronouncements contradict each other. The two foundation stones of modern physics, Relativity and quantum theory, appear to be telling us quite different things about the world.
\end{quote}
Essentially, we unify QM and SR by viewing SR as a principle explanation as above, then completing QRP via principle explanation with the same explanans, i.e., the relativity principle. 

What we mean by \textit{unification} here is simply ``to relate two or more disparate concepts or structures.'' For example, Maxwell unified the electric and magnetic fields by showing how the electric field can create the magnetic field and vice-versa. Then, Einstein unified Maxwell's electromagnetism and mechanics by showing how both share the same M4 kinematics of SR. This resolved the true conflict between Lorentz-invariant Maxwell's equations and Galilean-invariant Newtonian mechanics kinematically, since the Galilean-invariant spacetime of Newtonian mechanics follows from M4 in the limit $\frac{v}{c} \ll 1$. 

\begin{figure}
\begin{center}
\includegraphics [width = \textwidth]{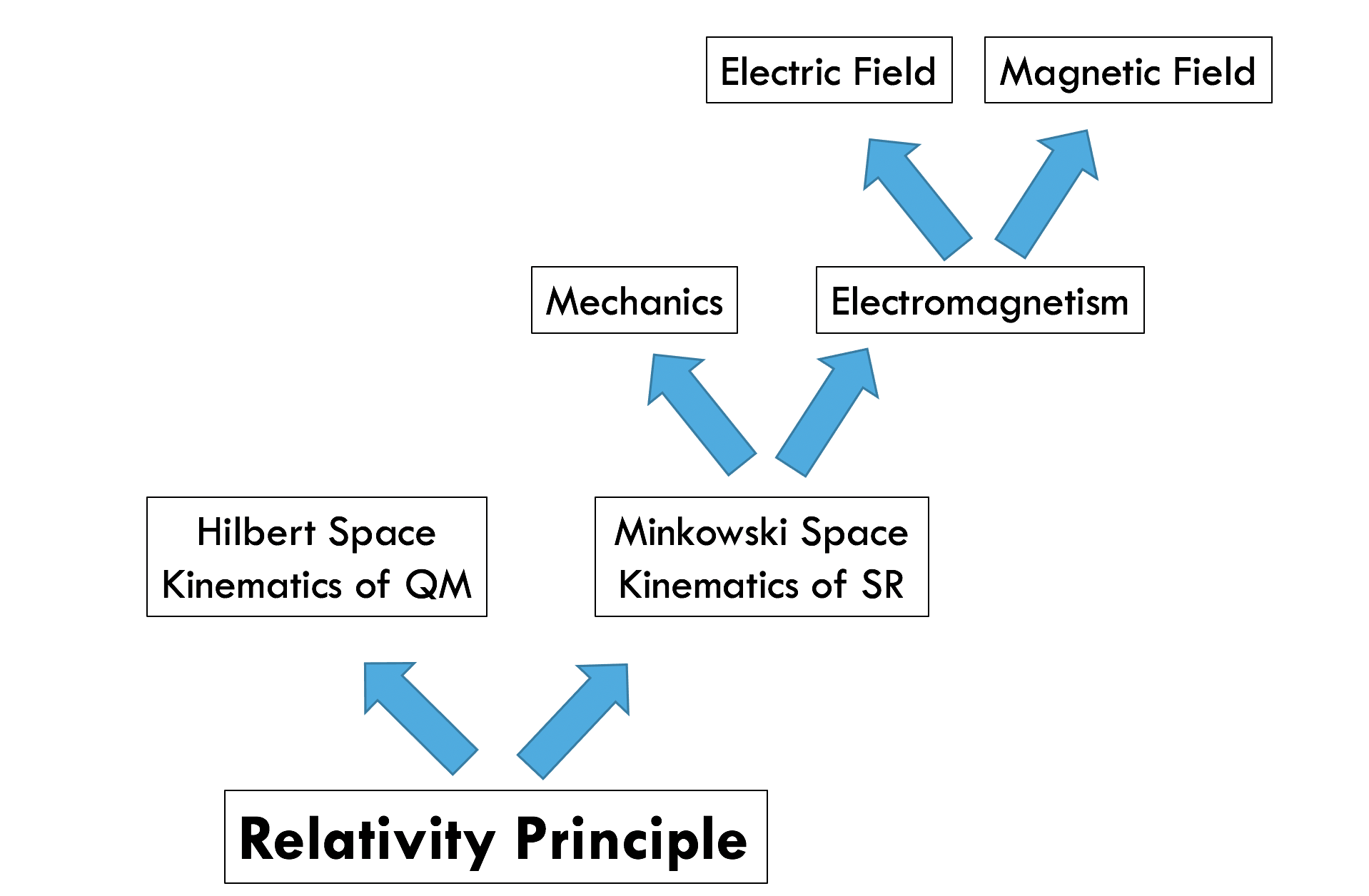}  
\caption{\textbf{Increased Unification}.} \label{Unification}
\end{center}
\end{figure}

In that spirit, we bring QM into the fold by showing how the disparate kinematic structures of SR (M4) and QM (Hilbert space) follow from the same relativity principle. So the historical pattern as shown in Figure \ref{Unification} is: two disparate concepts (electric and magnetic fields) unified by one theory (electromagnetism) followed by two disparate theories (electromagnetism and mechanics) unified by one kinematics (M4) followed by two disparate kinematics (M4 and Hilbert space) unified by one principle (relativity principle). 

Of course, we need to address the perceived conflict between QM and SR due to the perceived conflict between their kinematics, i.e., the entangled states of Hilbert space and the causal structure of M4. First, the correlations of entangled states are Lorentz invariant. Second, assuming those entangled states are complete, there is no `hidden' dynamics to account constructively for spacelike related quantum outcomes, so the entangled states per se cannot violate Lorentz covariance. This holds even though the time evolution of the state vector in Hilbert space (dynamics) is governed by Schr\"odinger's equation which is not Lorentz covariant, because we're only concerned with the distribution of quantum outcomes in M4. For example, one can use the path integral approach for computing these distributions and forego talk of the time evolution of the state vector altogether.  

In contrast, one might attempt to render SR a constructive explanation by replacing the relativity principle with a constructive account of the light postulate. For example, movement through the luminiferous aether causes length contraction in just the right way such that everyone measures the same value for the speed of light $c$, regardless of their relative motions. \cite[p. 221]{maudlinQNLandSR} writes:
\begin{quote}
    For example, it is possible to design theories that are empirically equivalent to the Special Theory of Relativity but that posit Newtonian Absolute Space and Absolute Time. If one supposes that Maxwell's equations hold in only the One True Reference Frame one can then derive that the behavior of electromagnetic clocks and measuring rods will not allow one to discover which inertial reference frame is the One True One. Rods will shrink and clocks will slow down in just such a way that the speed of light \textit{seems} to be the same in all frames, though it is not. 
\end{quote}

\clearpage

In that case, the derivation of the Lorentz transformations from the light postulate, i.e., SR as a principle theory, is unaffected while the relativity principle is arguably refuted by the preferred reference frame of the aether. However, the physics community long ago abandoned any widespread effort to render SR a constructive explanation while SR as a structural or principle explanation is widely accepted. \cite[p. 221]{maudlinQNLandSR} continues:
\begin{quote}
    Such a theory, although logically consistent and empirically impeccable, is generally considered to be inferior to Special Relativity. The grounds for this judgement are not usually made very explicit, but the general idea is that it would be awfully deceptive to create a world with Absolute Space and then use the laws of physics to hide its existence from us.
\end{quote}
Therefore, the principle theory of QM that QRP has produced might provide an understanding of QM as robust as our understanding of SR should it be completed as a principle explanation via the relativity principle. That is precisely what we will do in this paper.

Of course, this opens the door to a debate on whether or not principle explanation can be considered explanatory at all. One might open yet another door into the debate on causation. We are not going to proceed through either of those doors in this paper. For all practical purposes, you may believe as Einstein did that \cite{howardSEP}:
\begin{quote}
    Ultimate understanding requires a constructive theory, but often, says Einstein, progress in theory is impeded by premature attempts at developing constructive theories in the absence of sufficient constraints by means of which to narrow the range of possible constructive theories.
\end{quote}
In Section \ref{SectionExplanation}, we will simply argue a la \cite{fuchsQMasQI} that:
\begin{quote}
    Where present-day quantum-foundation studies have stagnated in the stream of history is not so unlike where the physics of length contraction and time dilation stood before Einstein's 1905 paper on special relativity.
\end{quote}
That is, the impasse that led Einstein to develop SR as a principle explanation as alluded to by \cite{howardSEP} was the failure by physicists to find a consensus constructive account of the light postulate, whence the Lorentz transformations. SR resulted when Einstein decided to abort ``constructive efforts'' and turned to principle explanation, justifying the light postulate with the relativity principle rather than deriving it via causal mechanisms. \cite{einsteinDespair} wrote:
\begin{quote}
    By and by I despaired of the possibility of discovering the true laws by means of constructive efforts based on known facts. The longer and the more despairingly I tried, the more I came to the conviction that only the discovery of a universal formal principle could lead us to assured results.
\end{quote}
Likewise, ``quantum-foundation studies have stagnated'' because of a pervasive failure of ``constructive efforts'' as noted by \cite{vancamp2011} above. 

Given this historical precedent, in what \cite{BerghoferIQOQI2023} describes as ``Perhaps the paper that can be regarded as the proper beginning of the quantum reconstruction project,'' \cite{rovelli1996} suggested we stop trying to \textit{interpret} QM and rather seek to \textit{derive} it in principle fashion:
\begin{quote}
    [Q]uantum mechanics will cease to look puzzling only when we will be able to \textit{derive} the formalism of the theory from a set of simple physical assertions (``postulates'', ``principles'') about the world. Therefore, we should not try to \textit{append} a reasonable interpretation to the quantum mechanics \textit{formalism}, but rather to derive the formalism from a set of experimentally motivated postulates.
\end{quote}
Again, \cite{rovelli1996} was motivated by the success of SR:
\begin{quote}
    The reasons for exploring such a strategy are illuminated by an obvious historical precedent: special relativity. ... Special relativity is a well understood physical theory, appropriately credited to Einstein’s 1905 celebrated paper. The formal content of special relativity, however, is coded into the Lorentz transformations, written by Lorentz, not by Einstein, and before 1905. So, what was Einstein’s contribution? It was to understand the physical meaning of the Lorentz transformations.
\end{quote}

The counterpart to the Lorentz transformations of SR is the Hilbert space of QM and Rovelli suggested ``simple physical assertions ... about the world'' might be found by viewing QM as a probability theory about information. This set in motion the axiomatic reconstruction of QM via information-theoretic principles. 

In Section \ref{SectionHistory}, we will (selectively) review the history of QRP. In short, \cite{hardy2001} produced the first of the so-called axiomatic reconstructions of QM based on information-theoretic principles with his 2001 paper, ``Quantum Theory from Five Reasonable Axioms'' \cite{jaeger2018}. Since then, many such axiomatic reconstructions of QM have been produced with what \cite{mueller2023} calls the ``first fully rigorous, complete reconstructions'' being produced by \cite{Chiribella2010} and \cite{masanesMuller2011}. What these reconstructions show (one way or another) is that the (finite-dimensional) Hilbert space formalism of QM can be derived from an empirically discovered fact called Information Invariance \& Continuity due to \cite{brukner2009}. Thus, QRP has succeeded in rendering QM a principle theory based on information-theoretic principles per their desideratum. 

\clearpage

Despite the longstanding success of QRP, the foundations community overall has not been sold on this information-theoretic understanding of QM. In Section \ref{SectionMissing}, we will explain two objections often voiced about QRP:
\begin{enumerate}
    \item The information-theoretic principles are ``highly abstract mathematical assumptions without an immediate physical meaning'' \cite{dakicBrukner2009}.
    \item The reconstructions do not contain anything beyond QM, so they do not ``offer more unification of the phenomena than quantum mechanics already does since they are equivalent'' \cite{vancamp2011}. 
\end{enumerate}
We will overcome these objections by proposing a completion of QRP via principle explanation a la SR. 

Specifically, in Section \ref{SectionPlanckPostulate} we explain how Information Invariance \& Continuity with spatialized measurement entails the \textit{Planck postulate}, i.e., everyone measures the same value for Planck's constant $h$, regardless of their relative spatial orientations or locations. Since spatial rotations and translations are part of the Poincar\'e group, and $h$ is a constant of Nature per Planck's radiation law, it's obvious that the Planck postulate can be justified by the relativity principle. In a sense, the Planck postulate is an ``interpretation'' of Information Invariance \& Continuity per the spatialization of QRP's operational notion of measurement along the lines called for by \cite{BerghoferIQOQI2023} and \cite{GoyalPhenomQBism2024}. 

Finally, since Information Invariance \& Continuity is (one way or another) at the foundation of QRP reconstructions of Hilbert space for QM, we have rendered QM a principle explanation (NPRF + $h$) with the same fundamental explanans as SR (NPRF + $c$). That is, the kinematics of QM and SR (Hilbert space and M4, respectively) are unified in that both follow from the relativity principle. Our understanding of QM is completed by realizing that Schr\"odinger's equation (dynamics) of QM, which gives the time evolution of the state vector in a fixed-dimensional Hilbert space, is simply the low-energy approximation of the Lorentz-invariant Klein-Gordon equation \cite[p. 172]{zee}. 

In Section \ref{SectionQuantClass} we show how the quantum-mechanical probabilities for the qubit and the joint probabilities for a Bell spin state follow from NPRF + $h$ \cite{NPRF2022}. We finish this section by pointing out that in the Stern-Gerlach measurement of spin, the polarization measurement of photons, and the double-slit experiment, NPRF + $h$ demands that a classically continuous quantity (angular momentum, energy, and momentum, respectively) be quantized such that the classically continuous prediction obtains on average over the distribution of quantum events. Thus, NPRF + $h$ can be understood as an adynamical global constraint over the distribution of quantum events in spacetime in accord with the ``all-at-once'' explanation of \cite{price,StuckeyFoP2008,EvansPriceWharton2010,EsfeldGisin2013,wharton3,ourbook,adlam2021,WhartonLiu2022,supermeasured2022,AdlamRovelli}. 

\clearpage

That is, per \cite{adlam2022} the mystery of entanglement isn't due to the individual properties of events in spacetime, but it's the result of some property of all the events involved. So, the constraints responsible for the correlations will be spatiotemporally global ``all-at-once'' constraints, which ``makes them a poor fit for a dynamic production picture.'' For example, in order for the Bell flash ontology of \cite{EsfeldGisin2013} to be relativistically covariant, you must consider entire possible histories or distributions of flashes in spacetime and give up stories about the temporal development of those histories or distributions. Likewise, \cite{AdlamRovelli} conclude that the ontology most compatible with relational quantum mechanics involves laws ``that apply atemporally to the whole of history, fixing the entire distribution of quantum events [throughout spacetime] all at once.''

In Section \ref{SectionConcl} we conclude that QM can be completed via principle explanation with ``an immediate physical meaning'' while unifying QM and SR in an unexpected way. Since you don't hear Nobel Laureates in Physics saying ``nobody understands special relativity,'' perhaps this completion of QM via the relativity principle will put an end to Nobel Laureates in Physics saying ``nobody understands quantum mechanics,'' regardless of whether or not a consensus constructive counterpart is ever found.


\section{Constructive, Structural and Principle Explanation}\label{SectionExplanation}

While both \cite{silberstein2021} and \cite{felline2011,felline2018SHPMP,Felline2018Synthese,Felline2021} (see also references therein) base constructive and principle explanation on Einstein's notions of constructive and principle theories (according to convention), the structural explanation of Felline differs importantly from the principle explanation of \cite{silberstein2021} used herein. Let's review constructive and principle theories per \cite{einstein1919}:
\begin{quote}
We can distinguish various kinds of theories in physics. Most of them are constructive. They attempt to build up a picture of the more complex phenomena out of the materials of a relatively simple formal scheme from which they start out. Thus the kinetic theory of gases seeks to reduce mechanical, thermal, and diffusional processes to movements of molecules – i.e., to build them up out of the hypothesis of molecular motion. When we say that we have succeeded in understanding a group of natural processes, we invariably mean that a constructive theory has been found which covers the processes in question.  \\

\noindent Along with this most important class of theories there exists a second, which I will call ``principle-theories.'' These employ the analytic, not the synthetic, method. The elements which form their basis and starting point are not hypothetically constructed but empirically discovered ones, general characteristics of natural processes, principles that give rise to mathematically formulated criteria which the separate processes or the theoretical representations of them have to satisfy. Thus the science of thermodynamics seeks by analytical means to deduce necessary conditions, which separate events have to satisfy, from the universally experienced fact that perpetual motion is impossible. \\

The advantages of the constructive theory are completeness, adaptability, and clearness, those of the principle theory are logical perfection and security of the foundations. The theory of relativity belongs to the latter class.
\end{quote}
Accordingly, \cite{silberstein2021} define constructive explanation in analogy with Einstein's definition of a constructive theory, so that causal mechanisms provide the explanans. This is consistent with the notion of mechanistic explanation per \cite{Felline2018Synthese} and mechanistic theory per \cite{maltranaEtAl2022}. Likewise, \cite{silberstein2021} define principle explanation in analogy with Einstein's definition of a principle theory, however they offer an important distinction from the structural explanation of Felline. 

That is, the explanans for principle explanation per \cite{silberstein2021} is a compelling fundamental principle to justify the empirically discovered fact at the foundation of a principle theory, while the explanans for structural explanation per Felline is the formal structure of the principle theory itself. Let's look at this distinction between principle explanation per \cite{silberstein2021} and structural explanation per Felline as it pertains to SR.


\subsection{Principle versus Structural Explanation}\label{SubSectPrinStructal}

We start with what is expected from a principle theory according to \cite{vancamp2011}:
\begin{quote}
If a principle theory approach is to succeed interpretationally, it must successfully play the explanatory role expected of a principle theory. It must establish the possibility of unification which gives a principle theory explanatory merit, or establish the conceptual framework necessary for clear empirical understanding. 
\end{quote}
Regarding ``the conceptual framework necessary for clear empirical understanding,'' Felline argues that length contraction is explained by the geometric structure of M4 without any underlying causal mechanism. \cite{Felline2021} writes:
\begin{quote}
     given the geometrical explanation showing how length contraction is the manifestation of a fundamental structure, there is no further, deeper causal/mechanistic/dynamical explanation to be found of this phenomenon. 
\end{quote}
And of course, SR as a principle theory provides unification by unifying mechanics and electrodynamics. Thus, this structural explanation satisfies Van Camp's explanatory requirements, regardless of whether one justifies the light postulate (whence the Lorentz transformations and M4) with the relativity principle or derives it by a causal mechanism a la the aether (although \cite{Felline2021} points out that an explanation via the formalism of a constructive theory is not a structural explanation). Not surprisingly, \cite[pp. 48-53]{maudlinQNLandSR} and many others are satisfied with this structural explanation, and we agree that it can be very useful. 

What then is the role of the relativity principle for this structural explanation? For those who want causation to be an essential element of explanation, the speed of light is ``the speed of causation'' \cite{PBSspeedOfLight} and the M4 structure of SR results from the invariance of the spacetime interval, which can be thought of as tracking ``causal proximity'' \cite{PBScausalM4}. This causal structure does not constitute a causal mechanism (otherwise, strictly speaking per \cite{Felline2021}, we don't have structural explanation), nonetheless it \textit{constrains} causal mechanisms, which is appealing to those with a constructive bias. In both cases, the relativity principle and light postulate are invoked separately towards that end. 

In contrast, \cite{Felline2018Synthese} avoids reference to causality altogether and uses the ``invariant hyperbolas'' of constant spacetime distance to the origin to explain length contraction geometrically. Of course, \cite{PBScausalM4} refers to the invariant hyperbolas in terms of ``communicating some causal influence.'' The point is, while causality is not necessary for structural explanation in SR, the formal structure of SR is certainly amenable to it and in that case, the relativity principle does play a supporting role.

In short, the relativity principle is not a primary explanans for structural explanation per Felline or \cite{Janssen2009} while it is precisely what elevates SR from a principle theory to a principle explanation per \cite{silberstein2021}. While both structural and principle explanation (rightly) have consensus support for SR, this distinction will prove to be important for unifying SR and QM via QRP. That's because structural explanation per QRP has not garnered much support and QRP does not even provide principle explanation (as defined here).


\subsection{Explanation in QRP}\label{SubSectionExplanRecon}

\cite{oddan2023} argues that QRP provides explanation via counterfactual dependence. For example, \cite{hardy2001} discovered that there is just one word (``continuous'') in one of his information-theoretic axioms that serves to distinguish classical probability theory from quantum probability theory (Section \ref{SectionHistory} below). \cite{oddan2023} notes that this is consistent with \cite{chiribella1} who believe ``We can only answer Wheeler's question `Why the quantum?' if we are able `to conceive of alternatives to quantum theory, ways the world \textit{might have been}'.'' However, even though there is consensus within the QRP community that the \textit{continuity} of reversible transformations between pure states is an important feature of QM, there are nonetheless many different viable reconstructions. This formal/operational fecundity in QRP is akin to the ontological fecundity in the interpretation program. In short, the QRP community is no closer to a long-sought consensus understanding of QM than the interpretation community, even if for very different reasons.

The reason structural explanation for QRP has failed to receive much support in the foundations community is that it doesn't satisfy either of Van Camp's explanatory requirements. For example, as we will see in Section \ref{SectionMissing}, the ``five simple physical requirements'' of \cite{masanesMuller2011} do not provide ``clear empirical understanding'' (unless you have a `physical intuition' for generalized probability theory) and they do not provide any unification beyond what is already contained in QM. Of course, as \cite{felline2018SHPMP} points out, there is some unification to be found in structural explanation per QM. 

For example, the Heisenberg Uncertainty Relation in its most general form applies to any pair of complementary variables, such as position and momentum or x-spin and z-spin. As \cite{felline2018SHPMP} notes:
\begin{quote}
    There is in fact no apparent sense in which the processes underlying the loss of a determinate position for a particle with definite momentum can be said to be the same as the one leading to the loss of x-spin for a particle with determinate z-spin. Such a relation is instead explained as part and parcel of the algebraic structure of observables in Quantum Theory.
\end{quote}
So, Van Camp's complaint about QRP is that it does not provide more unification than we already have in QM.

And, again, while the reconstructions have rendered QM a principle theory (finite-dimensional Hilbert space formalism of QM from an empirically discovered fact), Information Invariance \& Continuity is neither justified by a compelling fundamental principle nor is it shown to follow from a causal mechanism. So, the information-theoretic reconstruction of QM does not provide a principle or constructive explanation at all (as defined here).

In completing QRP via principle explanation with the relativity principle as the foundational explanans, we provide a ``clear empirical understanding'' of QM while unifying QM and SR, which are widely believed to be incompatible due to quantum entanglement (entanglement). This comes with an additional benefit, namely it provides a response to detractors of SR as a structural or principle explanation.

That is, some believe structural or principle explanation per SR lacks explanatory power, since it is without a constructive counterpart. Famous among these detractors is Brown. Concerning the principle explanation of length contraction, \cite{brownpooley2006} write:
\begin{quote}
    What has been shown is that rods and clocks must behave in quite particular ways in order for the two postulates to be true together. But this hardly amounts to an explanation of such behaviour.
\end{quote}
And, \cite{BrownTimpson2006} argue ``special relativity should not be a template for a fundamental reformulation of quantum mechanics.'' Brown and Timpson make their case using the information-theoretic reconstruction of QM by \cite{cliftonBubHalvorson2003} who write:
\begin{quote}
The foundational significance of our derivation, as we see it, is that quantum mechanics should be interpreted as a \textit{principle theory}, where the principles at issue are information-theoretic.
\end{quote}
Of course, as explained above, Brown and Timpson echo the consensus in foundations regarding the failure of QRP to support meaningful structural explanation and, as we have noted, QRP does not even attempt principle explanation. So, we concede this point. However, we do have a response to their argument as it pertains to SR, which they base on the constructive completion of thermodynamics. Our response concerns our distinction between structural and principle explanation while revealing the breadth of unification achieved by our proposed completion of QM.


\subsection{The Constructive Bias}\label{SubSectionConstBias}

Note that \cite{einstein1919} wrote, ``When we say that we have succeeded in understanding a group of natural processes, we invariably mean that a constructive theory has been found which covers the processes in question.'' So if the role of explanation is to provide understanding \cite{vancamp2011,Janssen2009}, clearly Einstein believed constructive explanation is fundamental to structural or principle explanation, i.e., constructive explanation is more basic or unifying than an alleged structural or principle explanation. For Einstein, the role of structural or principle explanation in the absence of a constructive explanation is to place constraints on any forthcoming constructive explanation, thereby narrowing the constructive possibilities. For instance, Einstein hoped that M4 would merely constrain some future constructive explanation for time dilation and length contraction. Here is what Einstein wrote to Arnold Sommerfeld in 1908 \cite{einsteinMichMorley1908}:
\begin{quote}
    It seems to me too that a physical theory can be satisfactory only when it builds up its structures from \textit{elementary} foundations. The theory of relativity is not more conclusively and absolutely satisfactory than, for example, classical thermodynamics was before Boltzmann had interpreted entropy as probability. If the Michelson-Morley experiment had not put us in the worst predicament, no one would have perceived the relativity theory as a (half) salvation. Besides, I believe that we are still far from satisfactory elementary foundations for electrical and mechanical processes. I have come to this pessimistic view mainly as a result of endless, vain efforts to interpret the second universal constant [$h$] in Planck's radiation law in an intuitive way.
\end{quote}
Notice specifically that Einstein believes mechanics and electrodynamics require a deeper unification than provided by SR. His standard for unification is the Boltzmann principle $S = k\ln{W}$, which he mentioned in his letter to Sommerfeld above. Einstein writes \cite{howardSEP}: 
\begin{quote}
This equation connects thermodynamics with the molecular theory. It yields, as well, the statistical probabilities of the states of systems for which we are not in a position to construct a molecular-theoretical model. To that extent, Boltzmann’s magnificent idea is of significance for theoretical physics ... because it provides a heuristic principle whose range extends beyond the domain of validity of molecular mechanics.
\end{quote}
So, while Einstein did have a clear preference for constructive explanation, he did appreciate any explanation that offered extensive unification. \cite[p. 32]{einsteinDespair} writes:
\begin{quote}
    A theory is the more impressive the greater the simplicity of its premises is, the more different kinds of things it relates, and the more extended its area of applicability.
\end{quote}
Note that in the same paragraph he complains about SR's lack of explanatory power, he laments ``vain efforts to interpret'' Planck's constant $h$. 

It may well be that Einstein hoped a constructive completion of QM, as necessitated by entanglement according to \cite{EPRpaper} (i.e., the Einstein-Podolsky-Rosen paradox), would produce an acceptable constructive counterpart to SR while providing a broader unification than SR alone (mechanics, electrodynamics, \textit{and} QM). This is certainly reasonable because QM is widely viewed as a theory of matter fundamental to classical physics and meter sticks and clocks are made of matter, so it would seem that length contraction and time dilation should be explicable most fundamentally by QM. 

But, a complete constructive account of QM a la \cite{EPRpaper} would have to violate locality, statistical independence, intersubjective agreement, and/or the uniqueness of experimental outcomes per Bell's theorem and the experimental violation of Bell inequalities \cite{Felline2021,silberstein2021,NPRF2024}. Of course, Einstein did not have the benefit of knowing about Bell's theorem, let alone the 2022 Nobel Prize in Physics awarded ``for experiments with entangled photons, establishing the violation of Bell inequalities.'' 

Armed with that knowledge, he might well have abandoned ``constructive efforts'' at understanding QM, viewing them as ``premature attempts at developing constructive theories in the absence of sufficient constraints,'' and again sought ``a universal formal principle [that] could lead us to assured results.'' In what follows, we show how QM completed via principle explanation per QRP plus the relativity principle, provides unification even broader than Einstein imagined. And in doing so, we undermine the complaint of those like \cite{BrownTimpson2006} that SR and QM should both be completed constructively, just as thermodynamics is completed constructively by statistical mechanics.


\subsection{Broad Unification: Countering the Constructive Bias}\label{SubSectionUnification}

Per Einstein, thermodynamics is a principle theory based on the empirically discovered fact that perpetual motion is impossible. For example, a perpetual motion machine of the second kind violates the second law of thermodynamics, i.e., entropy always increases. 

Specifically, the change in entropy $\Delta S_S$ for some system is equal to the amount of heat it gains $\Delta Q_S$ divided by its temperature $T_S$. The change in entropy for the environment of that system $\Delta S_E$ in this process would then be equal to the amount of heat it gains $\Delta Q_E$ divided by its temperature $T_E$. Since the amount of heat gained(lost) by the system equals the amount of heat lost(gained) by the environment, we have $\Delta Q_S = -\Delta Q_E$, so:
\begin{center}
$\Delta S_{Total} = \Delta S_S + \Delta S_E = \frac{\Delta Q_S}{T_S} - \frac{\Delta Q_S}{T_E} = \left(\frac{1}{T_S} - \frac{1}{T_E}\right)\Delta Q_S$
\end{center}
The second law says $\Delta S_{Total}$ must be greater than zero (it must increase), so if $\Delta Q_S$ is positive (the system gains heat from the environment), then it must be the case that $T_S < T_E$. Conversely, if $\Delta Q_S$ is negative (the system loses heat to the environment), then it must be the case that $T_E < T_S$. 

That is, heat must flow spontaneously from high temperature to low temperature. If someone designs a perpetual motion machine based on heat flowing spontaneously from low temperature to high temperature, then we can say it is ruled out by the second law of thermodynamics. Likewise, a perpetual motion machine of the first kind violates the first law of thermodynamics (conservation of energy) and a perpetual motion machine of the third kind violates the third law of thermodynamics (a heat engine cannot achieve 100\% efficiency). 

Conversely, one could say that thermodynamics is based on the empirically discovered fact that perpetual motion is impossible, so that:
\begin{center}
    \textbf{No perpetual motion of $X^{th}$ kind $\rightarrow X^{th}$ law of thermodynamics.}
\end{center}
That's the way Einstein viewed thermodynamics. \cite{BrownTimpson2006} point out that thermodynamics provides ``phenomenological laws which stipulate nothing about the deep structure of the working substance'' like we get from the constructive kinetic theory of matter. There we understand that the high temperature environment contains particles of higher kinetic energy on average than the particles of the lower temperature system. And, as the faster particles strike the slower particles, the slower particles speed up a bit while the faster particles slow down a bit. That's just in accord with conservation of momentum in a collision process and entails that heat flows from high temperature to lower temperature. So, the constructive account reverses the explanatory hierarchy of the principle account:
\begin{center}
    \textbf{Constructive kinetic theory of gases $\rightarrow X^{th}$ law of thermodynamics $\rightarrow$ No perpetual motion machines of the $X^{th}$ kind.}
\end{center}
Clearly, Brown and Timpson argue, the causal mechanisms of the constructive explanation provide a more compelling explanans than the empirically discovered fact (which provides neither a principle nor a structural explanation) and we would certainly agree. Continuing, \cite{BrownTimpson2006} argue:
\begin{quote}
    It is a remarkable thing that what might be called the kinematic structure of quantum theory, the nature of its observables and state space structure, can it seems be given a principle-theory, or `thermodynamic' underpinning. As Bell stressed, the beauty of thermodynamics is in its economy of reason, but the insight it provides is limited in relation to the messier story told in statistical mechanics.
\end{quote} 
If that was the end of the story, we would again agree with Brown and Timpson that QM as a principle theory based on the information-theoretic counterpart to complementarity/superposition/noncommutativity cries out for completion, just like thermodynamics based on the impossibility of perpetual motion. And, the compelling constructive completion of thermodynamics via statistical mechanics does suggest we look for a constructive completion of QM. However, the story doesn't end here for QM or thermodynamics. 

That is, by showing how the relativity principle justifies the empirically discovered fact at the foundation of information-theoretic reconstructions of QM, we see that QM provides a principle \textit{explanation} of entanglement as compelling as SR's principle explanation of length contraction. And, both theories are without a consensus constructive completion after many decades of effort. The addition of a compelling fundamental principle (relativity principle) that justifies the empirically discovered facts (Planck and light postulates) is what completes QM and SR as principle explanations. Otherwise, the fundamental explanans for length contraction and entanglement are the equally mysterious light and Planck postulates, respectively.

\begin{figure}
\begin{center}
\includegraphics [width=\textwidth]{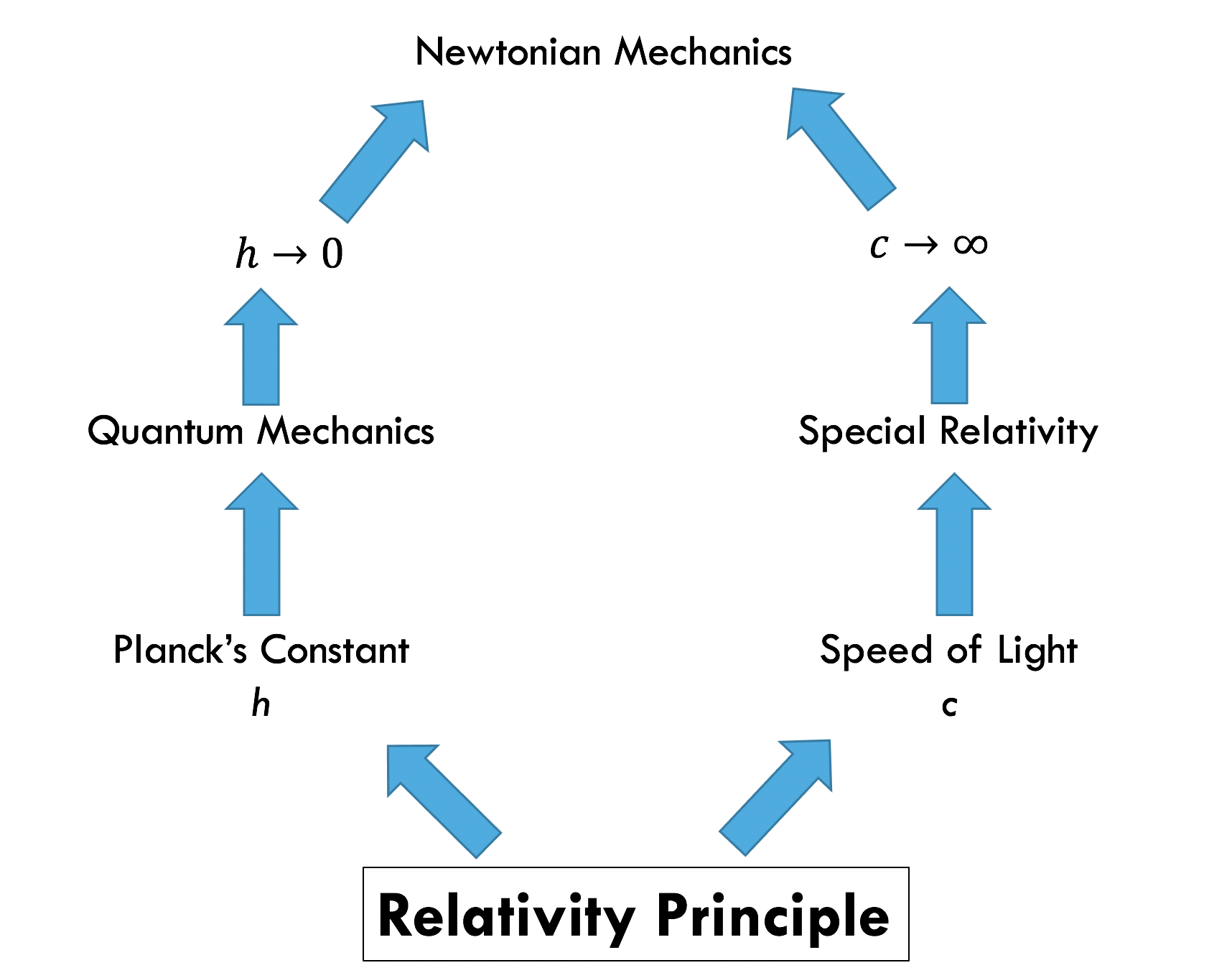}  \caption{The variables in Newtonian mechanics commute which means $h \rightarrow 0$ in the commutator for the corresponding variables in quantum mechanics. Newtonian equations hold on average according to quantum mechanics and follow from the corresponding equations in special relativity with $c \rightarrow \infty$.} \label{QM-SR-Newton}
\end{center}
\end{figure}

\begin{figure}
\begin{center}
\includegraphics [width = \textwidth]{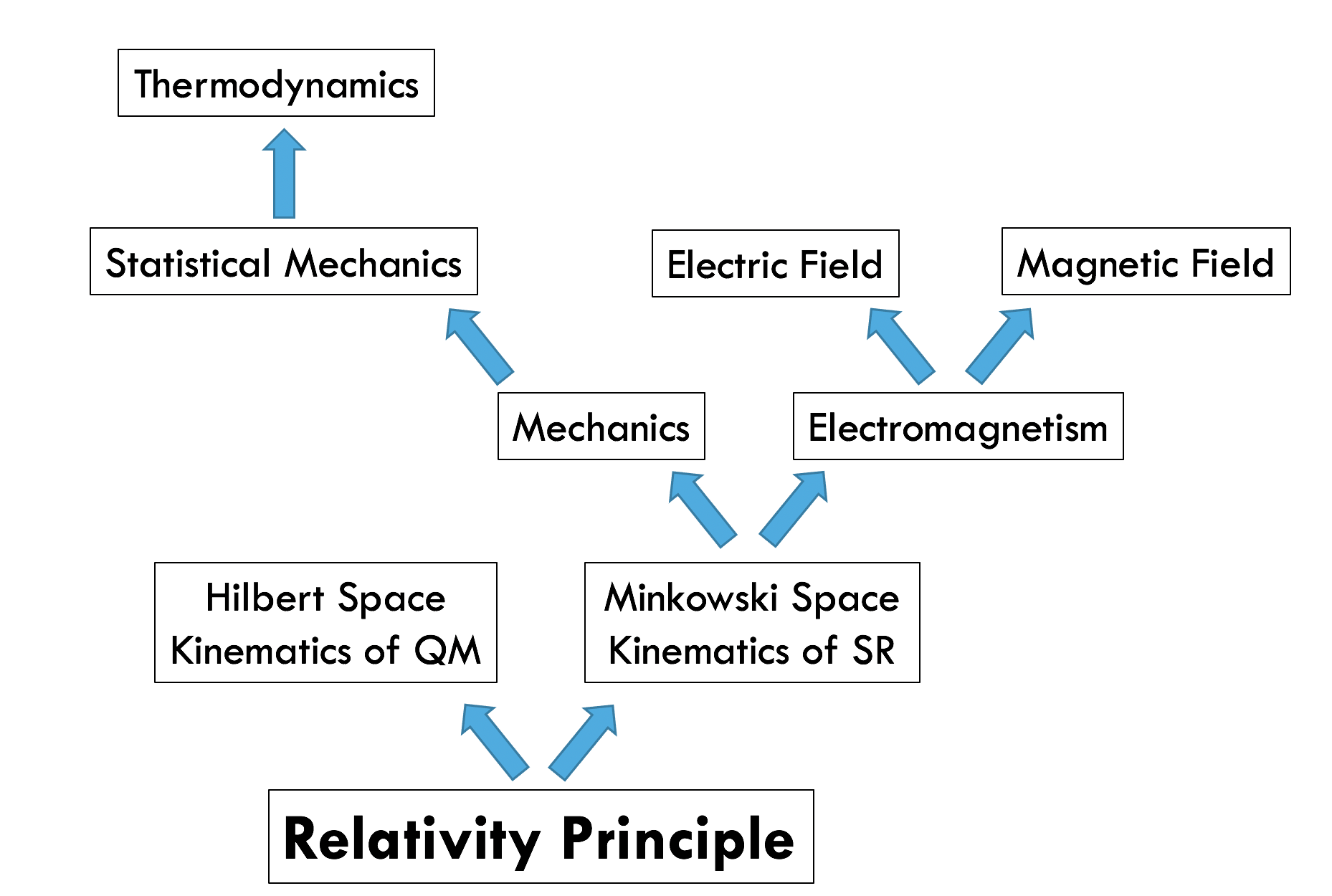}  
\caption{\textbf{Broad Unification}.} \label{UnificationBig}
\end{center}
\end{figure}

Likewise, thermodynamics as a principle theory based on the empirically discovered fact of ``no perpetual motion'' needs to be completed with a compelling fundamental principle to qualify as a principle explanation. Ironically, it finds that principle justification via its constructive completion. That is, since statistical mechanics uses relativistic mechanics or Newtonian mechanics, which is merely an approximation to QM and SR (Figure \ref{QM-SR-Newton}), we see that the most \textit{foundational} completion of thermodynamics isn't constructive, but principle. And to make matters worse for Brown and Timpson, that ultimate compelling fundamental principle is none other than the relativity principle (Figure \ref{UnificationBig}).  

\clearpage


\subsection{Summary}\label{SubSectionSummary}
 
The constructive bias is so strong that some philosophers, such as \cite{brownBook2005,brown2018}, have argued for our hypothetical defense of Einstein above. As \cite[pp. vii-viii]{brownBook2005} puts it:
\begin{quote}
    In a nutshell, the idea is to deny that the distinction Einstein made in his 1905 paper between the kinematical and dynamical parts of the discussion is a fundamental one, and to assert that relativistic phenomena like length contraction and time dilation are in the last analysis the result of structural properties of the quantum theory of matter.
\end{quote}
But, when one appreciates SR as a principle explanation over and above a structural explanation, it becomes clear that length contraction isn't about the mechanical or constructive shrinking of meter sticks. Length contraction is a relative, kinematic effect according to the relativity of simultaneity as necessitated by the relativity principle, whence the observer-independence of $c$. Likewise, when one truly appreciates QM as a principle explanation it becomes clear that entanglement isn't about the violation of locality, statistical independence, intersubjective agreement, or the uniqueness of experimental outcomes. Entanglement is a relative, kinematic effect according to `average-only' conservation as necessitated by the relativity principle, whence the observer-independence of $h$ (Sections \ref{SectionPlanckPostulate} and \ref{SectionQuantClass} below) \cite{stuckey2019,stuckey2020,NPRF2022,NPRF2024}. That means the most foundational explanans for thermodynamics is the relativity principle, not the causal mechanisms of statistical mechanics from mechanics (Figure \ref{UnificationBig}).

Pointing to an historical trend from the nineteenth century, \cite{Giovanelli2023} notes that:
\begin{quote}
    ... despite its apparent radical novelty, the relativity principle, like the energy principle, is ultimately an instance of ``that general direction of physical thought, which has been called the `physics of principles' in contrast to the physics of pictures and mechanical models'' (Cassirer, 1921b, 16; tr. 1923b, 359).
\end{quote}
And: 
\begin{quote}
The initial contradiction between mechanics and electrodynamics that is revealed by the negative result of ether drift experiments was overcome not ``by using the electrodynamic processes as a key to the mechanical'' but by establishing ``a far more perfect and deeper unity between the two than previously existed'' (Cassirer, 1921b, 33; tr. 1923b, 373). The unification of the two separate fields of theoretical physics -- electrodynamics and mechanics -- is not obtained through a process of horizontal integration, a reduction of the one to the other, but through a vertical integration, a subsumption of both theories under a higher principle.    
\end{quote}
So in summary, it appears to us that ``nobody understands quantum mechanics'' because `everyone' is wed to ``the physics of pictures and mechanical models'' used for ``deriving the principles from what are believed to be the laws of nature,'' rather than testing ``the acceptability of the laws of nature through certain general principles'' \cite{Giovanelli2023}. According to the strategy we are proposing here for understanding QM, to co-opt \cite{Giovanelli2023}:
\begin{quote}
    The unification of the two separate fields of theoretical physics -- SR and QM -- is not obtained through a process of horizontal integration, a reduction of SR to QM, but through a vertical integration, a subsumption of both theories under the relativity principle.
\end{quote}
And this is made possible by QRP, as we now explain.


\section{The Quantum Reconstruction Program}\label{SectionHistory}

\cite{GoyalPhenomQBism2024} points out that the idea of reconstructing a theory of physics traces back to classical mechanics, which was followed by Einstein's reconstruction of the Lorentz transformations via an operational framework, i.e., ``by positing a reasonable definition of light-based synchronization of distant clocks, and then showed how they led to the Lorentz transformations.'' Essentially, Einstein's reconstruction of the Lorentz transformations led to a new interpretation of Maxwell's electrodynamics. 

\cite{GoyalPhenomQBism2024} and \cite{jaeger2018} note that attempts to reconstruct abstract quantum formalism predates the reconstructions of interest here, i.e., those based on information-theoretic principles. \cite{GoyalPhenomQBism2024} writes:
\begin{quote}
    Recognition of the importance of reconstruction for elucidating the quantum formalism was not lost on the founders. For example, Heisenberg recognized that it would be highly desirable if the quantum formalism could somehow be derived using his uncertainty principle as a key axiom.
\end{quote}
However, the idea that QM deals fundamentally with information goes back at least to Bohr, as summed up by this famous 1958 statement by Petersen about Bohr's belief \cite{merminBohrQuote}:
\begin{quote}
    There is no quantum world. There is only an abstract quantum physical description. It is wrong to think that the task of physics is to find out how nature is. Physics concerns what we can say about Nature.
\end{quote}

\clearpage

As \cite{cavesFuchsSchack2001} note, this is particularly relevant ``for information-based interpretations of quantum mechanics, where quantum states, like probabilities, are taken to be states of knowledge rather than states of nature.'' And \cite{wheeler2}, who was greatly influenced by Bohr, wrote:
\begin{quote}
    No element in the description of physics shows itself as closer to primordial than the elementary quantum phenomenon, that is, the elementary device-intermediated act of posing a yes-no physical question and eliciting an answer or, in brief, the elementary act of observer-participancy. Otherwise stated, every physical quantity, every it, derives its ultimate significance from bits, binary yes-or-no indications, a conclusion which we epitomize in the phrase, it from bit.
\end{quote}
 This is Wheeler's famous ``It from Bit'' hypothesis, i.e., that physical objects (It) are based on information (Bit). Of course, QM might be dealing with information most fundamentally even if the converse of Wheeler's hypothesis is true, i.e., information requires physical objects to exist. For example, \cite{preskill2012} writes:
\begin{quote}
The moral we draw is that ``information is physical'' and it is instructive to consider what physics has to tell us about information. But fundamentally, the universe is quantum mechanical. How does quantum theory shed light on the nature of information?
\end{quote}
\cite{landauer1991} agrees stating:
\begin{quote}
    Information is not a disembodied abstract entity; it is always tied to a physical representation. 
\end{quote}

We don't need to commit either way on the ontology here, both extremes are compatible with our proposed completion of QRP (and therefore, QM). All that matters here is that QM can be understood as a theory about information. That fact and a particular complaint of Feynman (below) motivated \cite{zeilingerFoundPrin} to posit his Foundational Principle for QM (below). 

\cite{cliftonBubHalvorson2003} focused on the algebraic difference between classical and quantum possibility spaces (Boolean versus non-Boolean, respectively) per Heisenberg's commutative versus noncommutative algebra of observables (classical mechanics versus QM, respectively). \cite{bubpit2010,bubbook,bub2020} compared this difference with the difference between the geometry of Euclidean spacetime (Newtonian mechanics) and Minkowski spacetime (SR). As \cite{bub2012b} sums it up:
\begin{quote}
Hilbert space as a projective geometry (i.e., the subspace structure of Hilbert space) represents the structure of the space of possibilities and determines the kinematic part of quantum mechanics. ... The possibility space is a non-Boolean space in which there are built-in, structural probabilistic constraints on correlations between events (associated with the angles between the rays representing extremal events) -- just as in special relativity the geometry of Minkowski space-time represents spatio-temporal constraints on events. These are kinematic, i.e., pre-dynamic, objective probabilistic or information-theoretic constraints on events to which a quantum dynamics of matter and fields conforms, through its symmetries, just as the structure of Minkowski space-time imposes spatio-temporal kinematic constraints on events to which a relativistic dynamics conforms.
\end{quote}
\cite{rovelli1996} also focused on this difference in commutativity by noting that information gained in the measurement of some property of a quantum system is lost when subsequently measuring a noncommutative/non-Boolean complementary property of that system. In this paper, we will follow Hardy's approach with its important precursors, which returns us to Zeilinger's Foundational Principle as a response to \cite[p. 57]{feynmanQM}:
\begin{quote}
    It always bothers me that, according to the laws as we understand them today, it takes a computing machine an infinite number of logical operations to figure out what goes on in no matter how tiny a region of space and no matter how tiny a region of time, ... why should it take an infinite amount of logic to figure out what one tiny piece of space-time is going to do?
\end{quote}

Zeilinger actually wrote two different forms of his Foundational Principle as given in this single statement by \cite{jaeger2018}:
\begin{quote}
    ``An elementary system carries 1 bit of information,'' because ``an elementary system represents the truth value of one proposition.''
\end{quote}
In the abstract of \cite{zeilingerFoundPrin}, we see how his Foundational Principle leads to the mystery of entanglement:
\begin{quote}
    In contrast to the theories of relativity, quantum mechanics is not yet based on a generally accepted conceptual foundation. It is proposed here that the missing principle may be identified through the observation that all knowledge in physics has to be expressed in propositions and that therefore the most elementary system represents the truth value of one proposition, i.e., it carries just one bit of information. Therefore an elementary system can only give a definite result in one specific measurement. The irreducible randomness in other measurements is then a necessary consequence. For composite systems entanglement results if all possible information is exhausted in specifying joint properties of the constituents.
\end{quote}
\cite{bruknerZeil1999} expanded the Foundational Principle of QM to:
\begin{quote}
    The total information carried by the system is invariant under such transformation from one complete set of complementary variables to another.
\end{quote}

\clearpage

\noindent \cite{bruknerZeil2003} further clarified that with:
\begin{quote}
We show that if, in our description of Nature, we use one definite proposition per elementary constituent of Nature, some of the essential characteristics of quantum physics, such as the irreducible randomness of individual events, quantum complementary and entanglement, arise in a natural way. Then quantum physics is an elementary theory of information.
\end{quote}
One more statement in \cite{bruknerZeil2003}:
\begin{quote}
Thus, if we gradually change the orientation of the magnets in a set of Stern-Gerlach apparata defining a complete set of mutually complementary observables, a continuous change of the information vector will result.
\end{quote}
leads to Information Invariance \& Continuity per \cite{brukner2009}: 
\begin{quote}
\textbf{The total information of one bit is invariant under a continuous change between different complete sets of mutually complementary measurements.} 
\end{quote}
\cite{dakicBrukner2009} emphasize the importance of continuity in the third axiom of their reconstruction:
\begin{quote}
    (3) (Reversibility) Between any two pure states there exists a reversible transformation. If one requires the transformation from the last axiom to be continuous, one separates quantum theory from the classical probabilistic one.
\end{quote}
\cite{mueller2023} notes that Daki\'c and Brukner's reconstruction greatly influenced \cite{masanesMuller2011} who say this about the importance of continuity:
\begin{quote}
... if Requirement 4 is strengthened by imposing continuity of the reversible transformations, then [classical probability theory] is ruled out and [quantum theory] is the only theory satisfying the requirements. This strengthening can be justified by the continuity of time evolution of physical systems.
\end{quote}
Their wording is similar to \cite{hardy2001} who notes that if you delete just one word (``continuous'') from his Axiom 5, ``then we obtain classical probability theory instead'' of quantum probability theory. Finally, \cite{grinbaum2007} writes:
\begin{quote}
We see that various axiomatic systems for quantum theory contain, under one form or another, the assumption of continuity, and it is this assumption which is largely responsible for making things quantum.
\end{quote}

\clearpage

There is one more requirement needed to reproduce quantum probability theory, i.e., a way to stipulate how the rest of the probability space (or possibility/Hilbert space) can be constructed from the most fundamental (binary) quantum bit of information (qubit). This is the second axiom for \cite{dakicBrukner2009}:
\begin{quote}
    (2) (Locality) The state of a composite system is completely determined by local measurements on its subsystems and their correlations.
\end{quote}
\cite{ballEntanglement} points out that this simply means there is nothing ``hidden'' or ``missing'' from the quantum formalism. This additional mathematical requirement is analogous to assuming linearity in addition to the light postulate in order to derive the Lorentz transformations.


\section{What's the Problem?}\label{SectionMissing}

These reconstructions of QM based on information-theoretic principles have clearly not won consensus support in the foundations community. One problem was noted by \cite{dakicBrukner2009}:
\begin{quote}
The vast majority of attempts to find physical principles behind quantum theory either fail to single out the theory uniquely or are based on highly abstract mathematical assumptions without an immediate physical meaning (e.g. [18]). ...\\

While [the operational] reconstructions are based on a short set of simple axioms, they still partially use mathematical language in their formulation. 
\end{quote} 
For example, here are the ``five simple physical requirements'' of \cite{masanesMuller2011}:
\begin{enumerate}
    \item In systems that carry one bit of information, each state is characterized by a finite set of
outcome probabilities.
\item The state of a composite system is characterized by the statistics of measurements on the
individual components.
\item All systems that effectively carry the same amount of information have equivalent state
spaces.
\item Any pure state of a system can be reversibly transformed into any other.
\item In systems that carry one bit of information, all mathematically well-defined measurements
are allowed by the theory.
\end{enumerate}
These establish classical probability theory and quantum probability theory uniquely among all generalized probability theories. As stated above, if you change Requirement 4 to read, ``Any pure state of a system can be \textit{continuously} reversibly transformed into any other,'' then you select quantum probability theory alone. This is the kinematics of QM, to get the dynamics (measurement update and Schr\"odinger's equation) \cite{masanesMuller2011} add two more requirements:
\begin{itemize}
    \item If a system is measured twice ``in rapid succession'' with the same measurement, the same outcome is obtained both times.
    \item Closed systems evolve reversibly and continuously in time.
\end{itemize}
Masanes and M\"uller's reconstruction is essentially Hardy's reconstruction minus his (unnecessary) Simplicity axiom \cite{mueller2023}. The problem is obvious -- unless you possess a `physical intuition' for generalized probability theories, these requirements are not likely going to tell you much about physical reality. Essentially, this is the point made by \cite{BerghoferIQOQI2023} that QRP needs an interpretation. \cite{GoyalPhenomQBism2024} also points out that the elucidation of QM via its reconstruction is a two-step process, i.e., ``Reconstruct the quantum formalism'' then ``Interpret the reconstruction,'' and admits that work remains for the second step. 

Another common complaint from the foundations community about QRP is noted here by \cite{vancamp2011}:
\begin{quote}
However, nothing additional has been shown to be incorporated into an information-theoretic reformulation of quantum mechanics beyond what is contained in quantum mechanics itself. It is hard to see how it could offer more unification of the phenomena than quantum mechanics already does since they are equivalent, and so it is not offering any explanatory value on this front.
\end{quote} 
These two problems can be solved by interpreting (and thereby completing) QRP via a compelling fundamental principle justifying Information Invariance \& Continuity (Planck postulate upon interpretation). Foundations in general is not going to accept that the noncommutativity/superposition/complementarity of the qubit solves the mystery of entanglement any more than they would accept the light postulate as a solution to the mystery of length contraction. The reason we can accept that the light postulate solves the mystery of length contraction is because the observer-independence of $c$ is a necessary consequence of the relativity principle and the relativity principle is compelling enough to provide a foundational explanans. 

That's why \cite{hohn2023} missed the analogy completely when he said, ``Entanglement from complementarity is not as intuitive as the relativity of simultaneity from the relativity principle.'' The proper analogy is that \textit{both} the relativity of simultaneity \textit{and} entanglement follow most fundamentally from the relativity principle. Likewise, a statement from M\"uller's Group at the Institute for Quantum Optics and Quantum Information (IQOQI) misses the analogy and fails to capture the true accomplishment of QRP. A corrected version of the statement would read:
\begin{quote}
Can quantum theory be derived from simple principles, in a similar way as the Lorentz transformations can be derived from \st{the relativity principle and} the constancy of the speed of light? The exciting answer is ``yes''! 
\end{quote}
 In other words, QRP has succeeded in rendering QM a principle theory, i.e., a mathematical formalism (Hilbert space of QM) derived from an empirically discovered fact (Information Invariance \& Continuity). What they don't have, as indicated by the struck through text in the corrected statement, is QM as a principle \textit{explanation}. Again, the difference between a principle theory and a principle explanation is that a principle explanation includes a compelling fundamental principle justifying the empirically discovered fact at the foundation of the principle theory, e.g., the relativity principle justifying the light postulate. 

Alternatively, one could propose a constructive account of the empirically discovered fact at the foundation of the principle theory, e.g., length contraction due to movement through the aether causing the observer-independence of $c$. No consensus aether account has ever been rendered for the light postulate while its justification via the relativity principle is so widely accepted that it is presented in introductory physics textbooks. Likewise, QRP has derived the Hilbert space of QM from Information Invariance \& Continuity, but has not provided a constructive account of Information Invariance \& Continuity or a compelling fundamental principle to justify it. Essentially, the constructive interpretations of QM are the `aether theories' for Information Invariance \& Continuity in this analogy and Bell's theorem with the violation of Bell inequalities tells us why they have not received (nor will they likely ever receive) consensus support \cite{silberstein2021,Felline2021, NPRF2024}. \cite{bub2004} writes:
\begin{quote}
    That is, just as Einstein's analysis (based on the assumption that we live in a world in which natural processes are subject to certain constraints specified by the principles of special relativity) shows that we do not need the mechanical structures in Lorentz's theory (the aether, and the behaviour of electrons in the aether) to explain electromagnetic phenomena, so the Clifton-Bub-Halvorson analysis (based on the assumption that we live in a world in which there are certain constraints on the acquisition, representation, and communication of information) shows that we do not need the mechanical structures in Bohm's theory (the guiding field, the behaviour of particles in the guiding field) to explain quantum phenomena. 
\end{quote}

\clearpage

What we will show in Section \ref{SectionPlanckPostulate} is that the principle of Information Invariance \& Continuity residing at the foundation of information-theoretic reconstructions (one way or another) represents an empirically discovered fact, i.e., the observer-independence of Planck's constant $h$ between different inertial reference frames, upon spatialization of the purely operational notion of measurement in QRP. The direct mathematical consequence of this is noncommutativity/superposition/complementarity, i.e., the `weirdness' of the qubit upon which the kinematic structure of QM is built in QRP.


\section{The Planck Postulate}\label{SectionPlanckPostulate}

So, a fundamental principle for the information-theoretic reconstructions of QM is Information Invariance \& Continuity. According to this principle, there exists a fundamental unit of information (the qubit) that represents a definite outcome with respect to only one measurement. 

\begin{figure}
\begin{center}
\includegraphics [width=\textwidth]{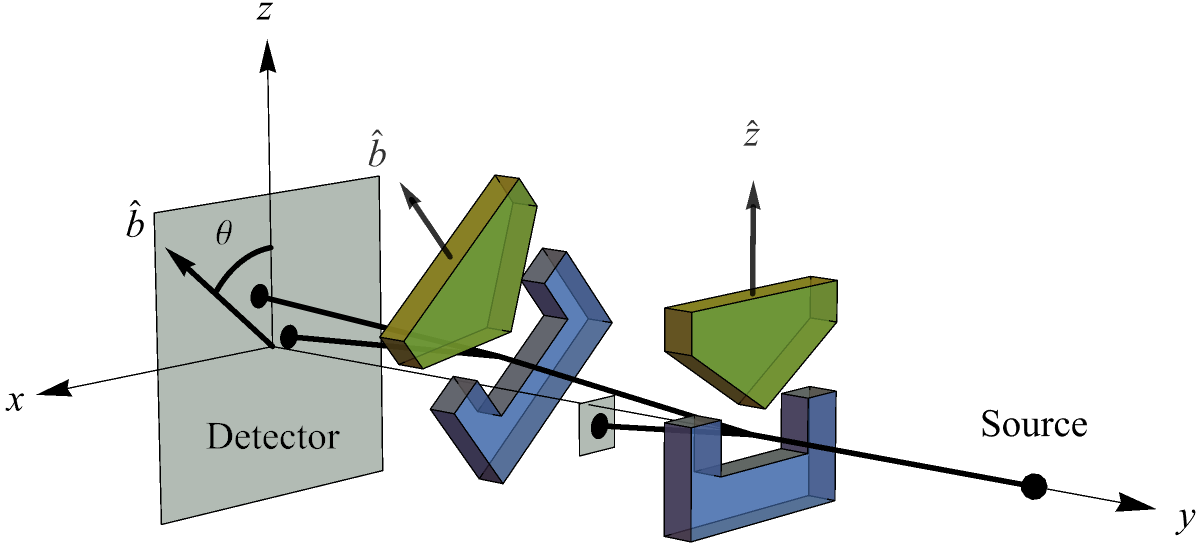}  \caption{In this set up, the first Stern-Gerlach (SG) magnets (oriented at $\hat{z}$) are being used to produce an initial state $|\psi\rangle = |z+\rangle$ for measurement by the second SG magnets (oriented at $\hat{b}$).} \label{SGExp1}
\end{center}
\end{figure}

\begin{figure}
\begin{center}
\includegraphics [height = 65mm]{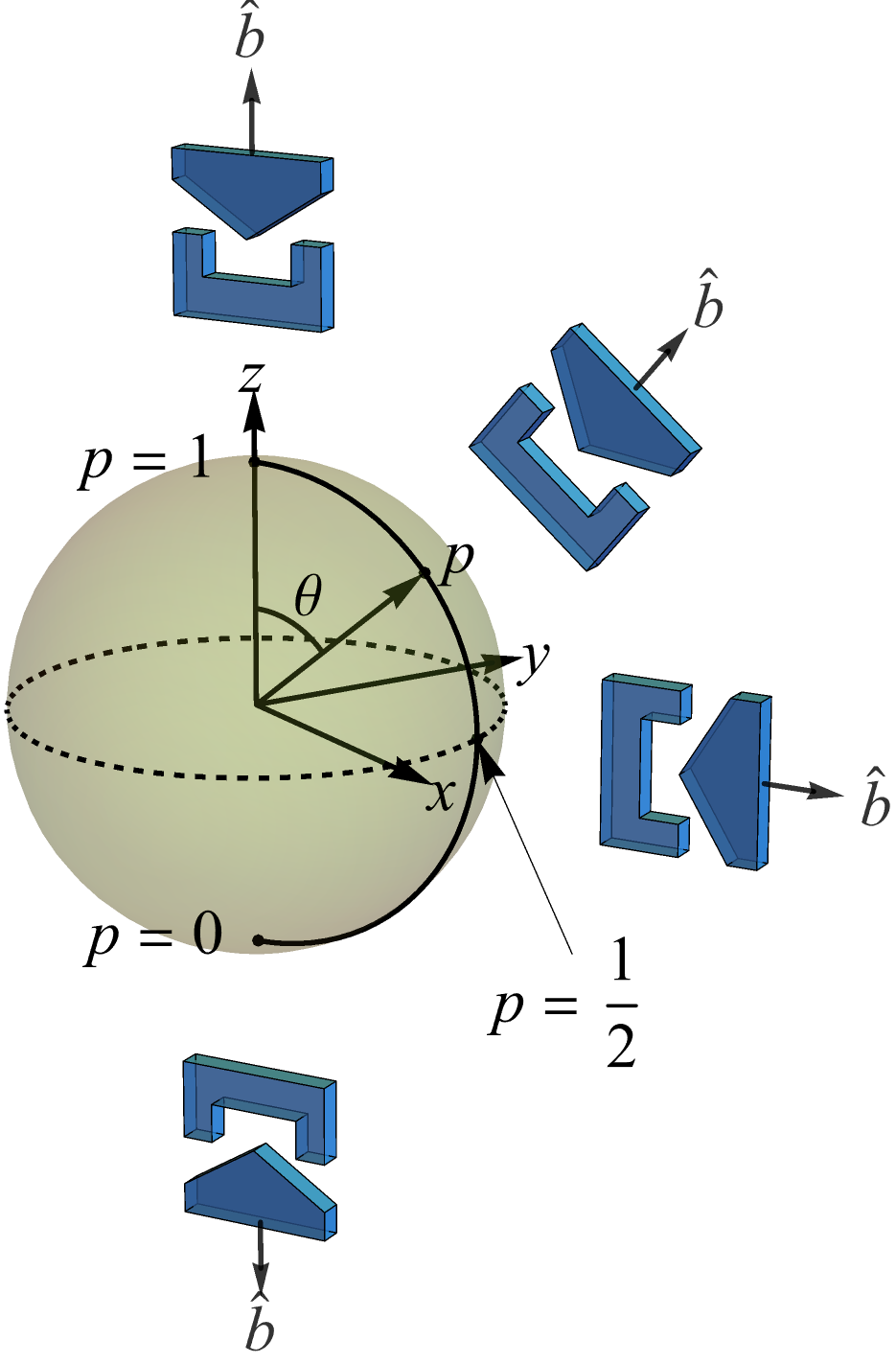}
\caption{Probability state space for the measurement of Figure \ref{SGExp1}. Since this state space is isomorphic to 3-dimensional real space, the Bloch sphere is shown in a real space reference frame with its related SG magnet orientations \cite{bruknerZeil2003}.} \label{Qubit}
\end{center}
\end{figure}

For example, the information contained in the qubit of Figures \ref{SGExp1} and \ref{Qubit} is simply ``$+1$ is the outcome of a Stern-Gerlach (SG) spin measurement in the $\hat{z}$ direction.'' Now suppose you ask, ``Is $+1$ the outcome of an SG spin measurement in the $\hat{b} \neq \hat{z}$ direction?'' The answer is indefinite, i.e., probabilistic. This is what \cite{zeilingerFoundPrin} meant when he wrote, ``an elementary system can only give a definite result in one specific measurement. The irreducible randomness in other measurements is then a necessary consequence.'' 

\begin{figure}
\begin{center}
\includegraphics [height = 75mm]{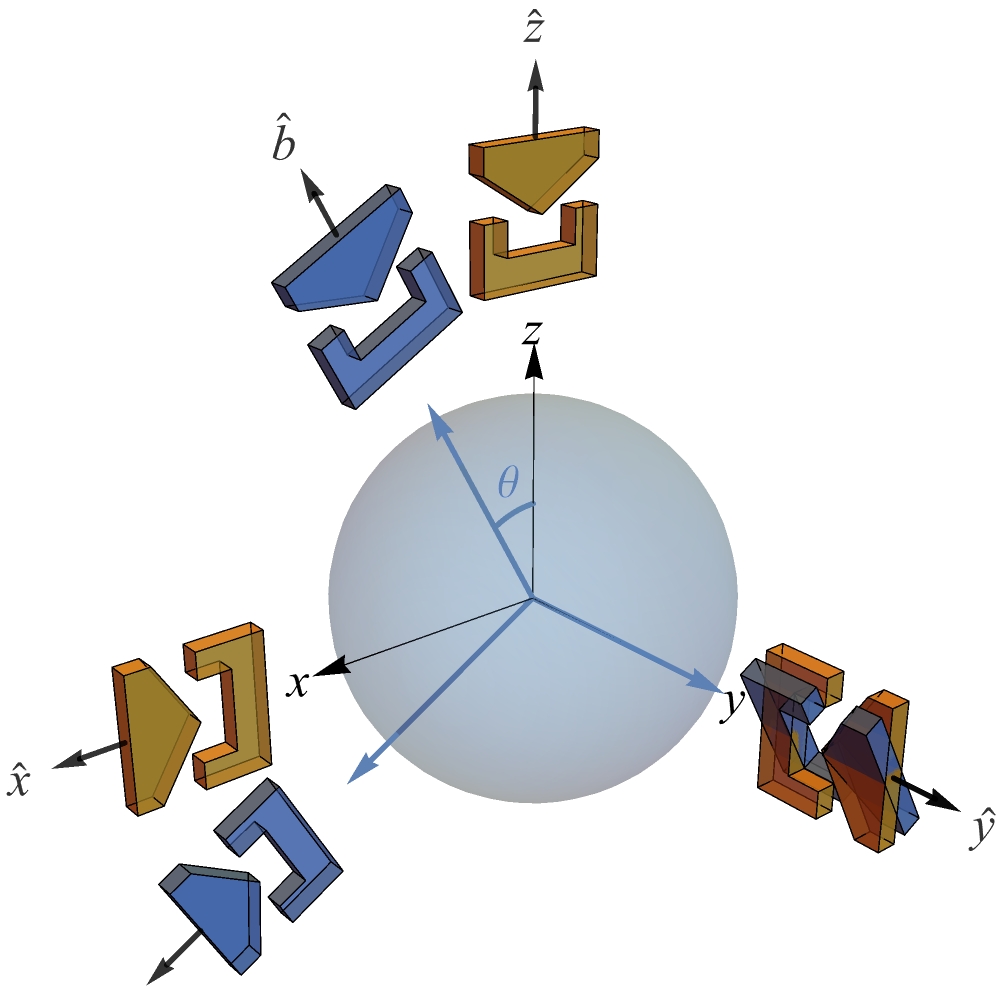} 
\caption{Reference frames for the complementary SG spin measurements associated with Figures \ref{SGExp1} and \ref{Qubit} \cite{bruknerZeil2003}.} \label{ComplBases}
\end{center}
\end{figure}

\begin{figure}
\begin{center}
\includegraphics [height = 55mm]{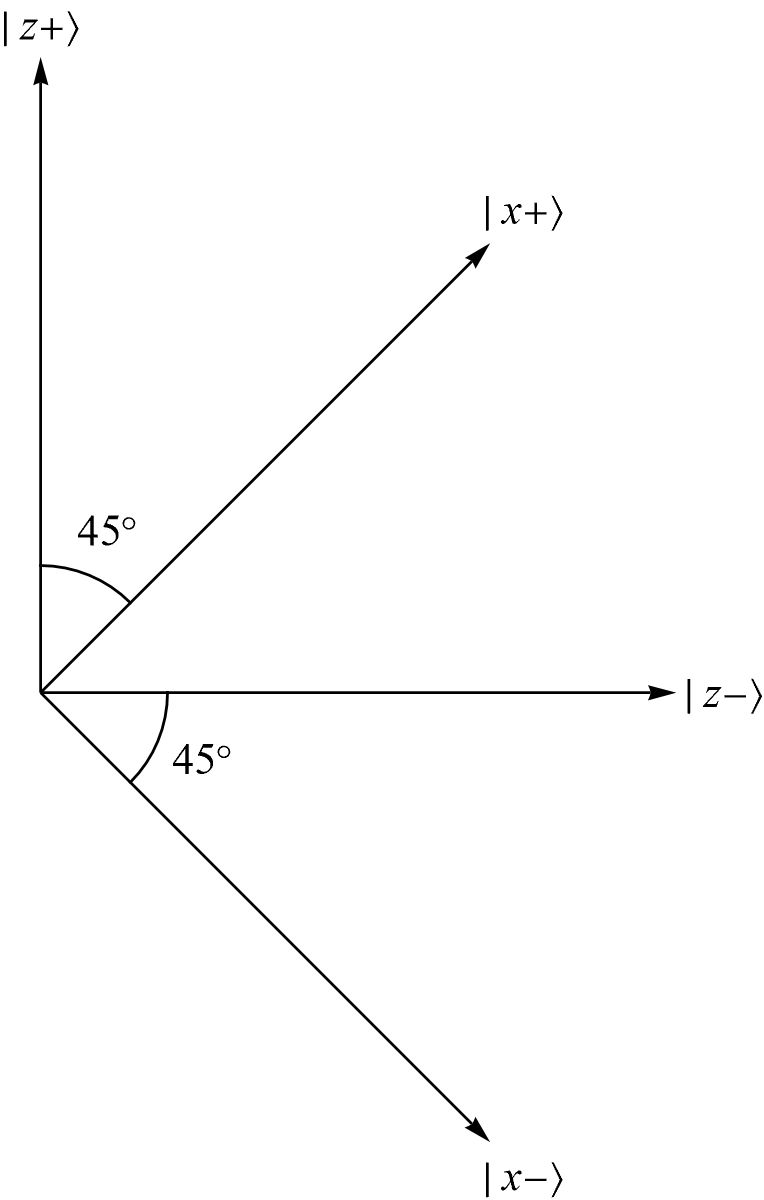}  \caption{The basis vectors of Hilbert space represent the possible outcomes of a measurement. Here we see the basis vectors for an SG spin measurement along $\hat{z}$ ($|z+\rangle$ representing a $+1$ outcome and $|z-\rangle$ representing a $-1$ outcome) and the basis vectors for an SG spin measurement along $\hat{x}$.} \label{HilbertSpace}
\end{center}
\end{figure}

Even though the results of other measurements are indefinite, the total information of the qubit is invariant. That is, information for a qubit is represented by a unit length vector in Hilbert space and that information is distributed by projection among the binary outcomes for other qubits, e.g., for SG spin measurements in the $xz$ plane (Figure \ref{ComplBases}) we have $\frac{|z+\rangle + |z-\rangle}{\sqrt{2}} = |x+\rangle$  (Figure \ref{HilbertSpace}). An orthogonal pair of Hilbert space vectors corresponds to opposing probability vectors in Figure \ref{Qubit}, e.g., $\theta = 0$ and $\theta = 180^{\circ}$ correspond to $|z+\rangle$ and $|z-\rangle$, respectively. So, superposition is associated with the continuous distribution of pure states for the Bloch sphere. 

\clearpage

And, as we said above, what makes the qubit different than a classical bit is that these different possible measurements are all related in continuous fashion. A classical bit has discrete measurement options, e.g., opening box 1 or box 2, each with two possible outcomes, e.g., find a ball or no ball. A qubit has continuous measurement options each with two possible outcomes. As \cite{darrigol2022} says, the kinematic (Hilbert space) structure of QM:
\begin{quote}
    results from the harmonious blending of the discontinuity of measurement results with the continuity of the possibilities of measurement.
\end{quote}
For the SG spin measurements, the continuous measurements are the orientations of the SG magnets and the two outcomes are ``up'' or ``down'' relative to the N-S direction of those magnets. For polarizers, the continuous measurements are the orientations of the polarizing axis and the two outcomes are ``pass'' or ``no pass'' for the photons incident on the polarizer. For the double-slit experiment, the continuous measurements are locations of the detector screen relative to the slits along the optic axis and the two possible outcomes are ``slit 1'' or ``slit 2'' for a position measurement and ``constructive'' or ``destructive'' interference for a momentum measurement \cite{NPRF2024}. 

Per \cite{GoyalPhenomQBism2024}, the spatial notions of measurement in these examples are not inherent in the operational framework of QRP:
\begin{quote}
    For example, we tend to think of an agent as an embodied being localized in space; a physical system as an object that is spatially localized in our laboratory at all times; or a measurement as carried out by a chunk of equipment in one corner of a laboratory. But the operational framework abstracts away all of these \textit{spatial} notions. So, a \textit{physical system} is simply an entity that \textit{persists} -- it does not necessary exist anywhere in particular at a given moment in time. A \textit{measurement} is an abstract parameterized process that acts on a physical system to generate an \textit{outcome} and to output the same physical system -- it is not a spatially localized piece of equipment. The agent is simply \textit{an entity} that exists and persists over time, and is capable of \textit{observing outcomes} and of \textit{freely acting} to change settings associated with measurement and interaction devices -- it is not a spatially localized human being.
\end{quote}
In that sense, the Planck postulate may be viewed as a \textit{spatial interpretation} of Information Invariance \& Continuity, constituting the second step in Goyal's ``elucidative strategy'' for quantum theory. 

To complete the interpretation of Information Invariance \& Continuity, we note that each measurement is associated with a reference frame per its complementary measurements and these reference frames are related by spatial rotations or translations. The complementary spin measurements, e.g., $S_x$ and $S_z$, are related by spatial rotations (Figure \ref{ComplBases}) as are the complementary polarization measurements. In the double-slit experiment, the complementary measurement configurations of position and momentum are related by spatial translations. 

Finally, all of this can be associated directly with Planck's constant $h$ since, as \cite{hohn2018} notes, $h$ represents ``a universal limit on how much simultaneous information is accessible to an observer.'' For example, for complementary spin measurements $(S_x,S_y,S_z)$ the commutator is $S_xS_y - S_yS_x=\textbf{i}\hbar S_z$ (which also applies to photon polarization \cite{Kim2010}). For the complementary measurements of position $x$ and momentum $p$ in the double-slit experiment the commutator is $xp - px = \textbf{i}\hbar$. And again, these complementary measurement configurations establish a reference frame related to other complementary measurement reference frames in continuous fashion via spatial rotations or spatial translations. Therefore, the invariance of the total information between these different reference frames means $h$ is the same in reference frames related by spatial rotations and translations, i.e., we have the \textit{Planck postulate} in analogy with the light postulate.

Putting all of this together we see that Information Invariance \& Continuity at the foundation of axiomatic reconstructions of QM is the information-theoretic counterpart to the conventional quantum characteristics of noncommutativity, superposition and complementarity. And upon spatialization of QRP's operational notion of measurement, it entails the invariance of $h$ per the measurement outcomes in inertial reference frames of different complete sets of mutually complementary measurements, which can obviously be justified by the relativity principle (NPRF + $h$).


\section{The Quantum-Classical Relationship}\label{SectionQuantClass}

\begin{figure}
\begin{center}
\includegraphics [height = 75mm]{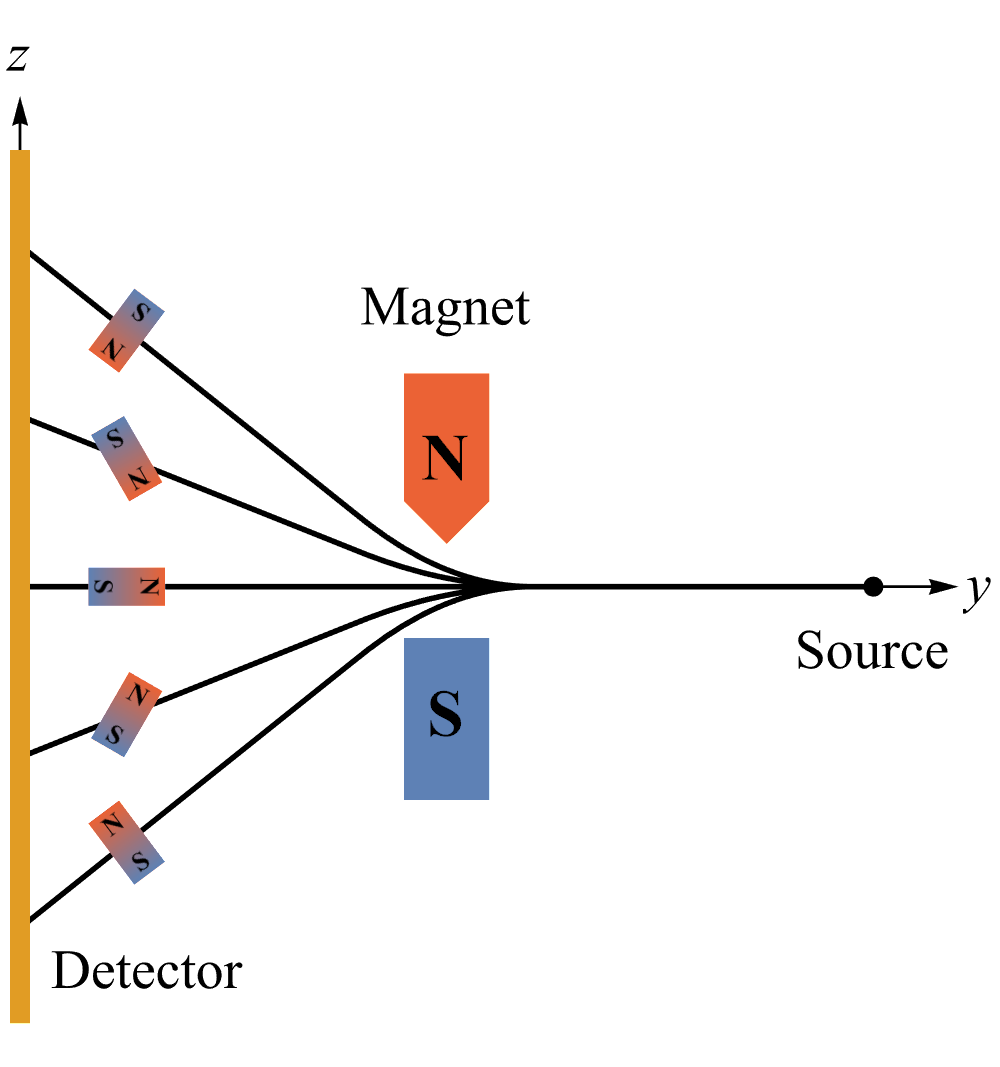}
\caption{The classical constructive model of the SG experiment. If the atoms enter with random orientations of their ``intrinsic'' magnetic moments (due to their ``intrinsic'' angular momenta), the SG magnets should produce all possible deflections, not just the two that are observed \cite{knight,franklin2019}. Compare with Figure \ref{SGExp2}.} \label{SGclassical}
\end{center}
\end{figure}

\begin{figure}
\begin{center}
\includegraphics [width=\textwidth]{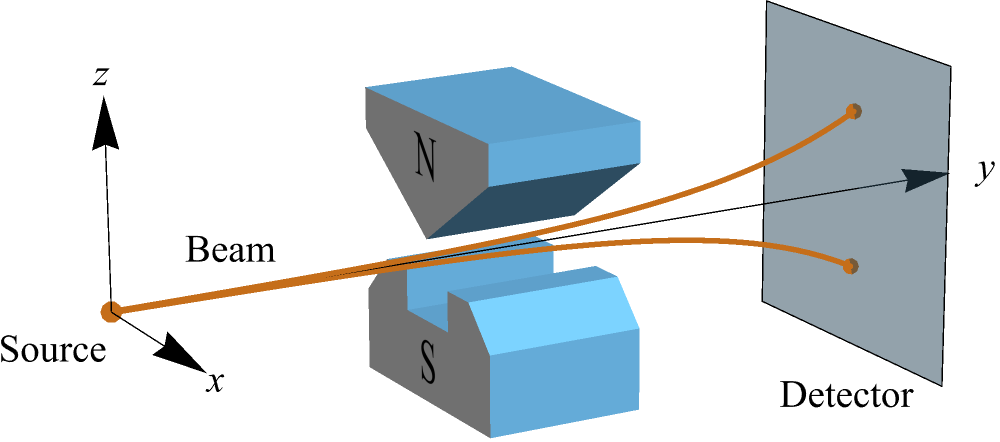}  \caption{An SG spin measurement showing the two possible outcomes, up ($+\frac{\hbar}{2}$) and down ($-\frac{\hbar}{2}$) or $+1$ and $-1$, for short. The important point to note here is that the classical analysis (Figure \ref{SGclassical}) predicts all possible deflections, not just the two that are observed.}  \label{SGExp2}
\end{center}
\end{figure}

\begin{figure}
\begin{center}
\includegraphics [height = 65mm]{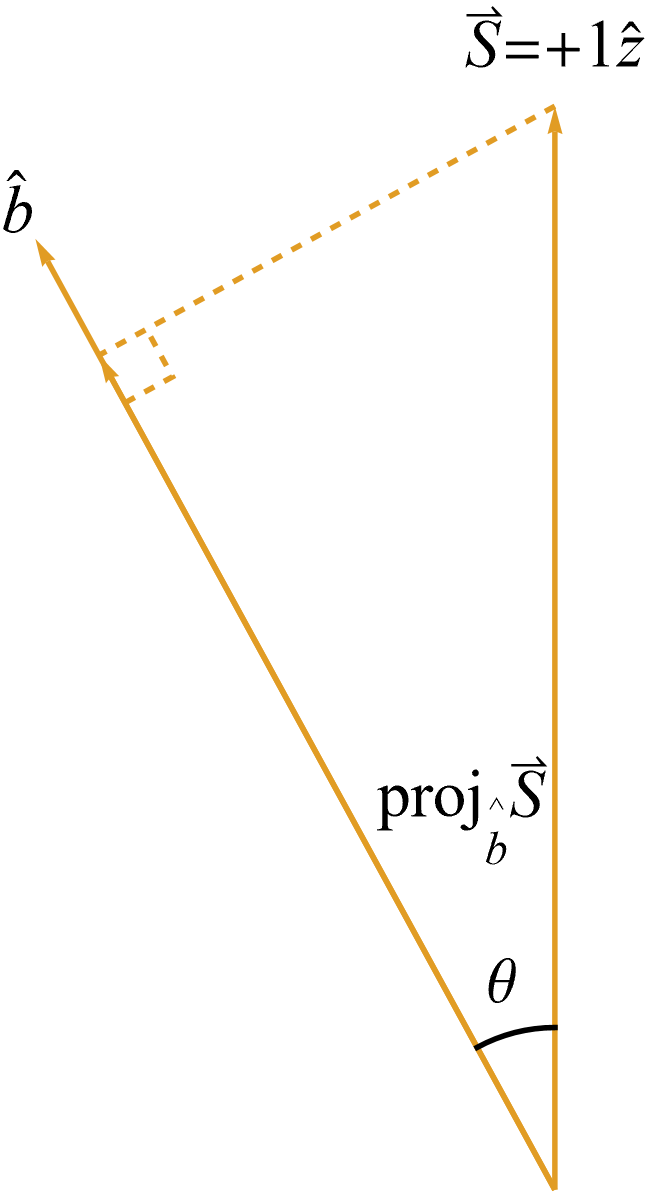} 
\caption{Spin angular momentum $\vec{S}$ along $\hat{z}$ projected along the direction $\hat{b}$. This does \textit{not} happen with spin angular momentum due to the relativity principle.} \label{Projection}
\end{center}
\end{figure}

If $h = 0$, then the complementary measurements would commute (they wouldn't be complementary) and we would have the classical situation instead of the quantum situation. For SG spin measurements, that would mean we have Figure \ref{SGclassical} instead of Figure \ref{SGExp2}. So, the measurement of $|z+\rangle$ along $\hat{x}$ or $\hat{y}$ would produce the projection of $+1\hat{z}$ along $\hat{x}$ or $\hat{y}$ as shown in Figure \ref{Projection}, which for $\hat{b}$ equal to $\hat{x}$ or $\hat{y}$ would be zero. In other words, given that we know the outcome of a $\hat{z}$ measurement is definitely going to be $+1$, we would also know the $\hat{x}$ and $\hat{y}$ measurement outcomes will be zero, i.e., simultaneous information is now available. But if that happened, our measurements would only be producing $h$ in the $\hat{z}$ direction while measurements in all other directions would be producing a fraction of $h$ because, as \cite{weinberg2017} points out, measuring an electron's spin via SG magnets constitutes the measurement of ``a universal constant of nature, Planck's constant $h$'' (Figure \ref{SGExp2}).

That is, Information Invariance \& Continuity with spatialized measurement entails that everyone must measure the same value for Planck's constant $h$, regardless of their SG magnet orientations relative to the source, which like the light postulate is an empirically discovered fact. That means the spin-$\frac{1}{2}$ particles must always be deflected the same amount up or down relative to the magnetic field of the SG magnets, regardless of their orientation relative to the source. Again, this prediction differs from the classical constructive model for spin-$\frac{1}{2}$ particles as shown in Figure \ref{SGclassical}.

A partial deflection in the SG experiment (fractional value of $h$) is totally analogous to measuring the speed of light and obtaining a fractional value of $c$. The understanding would be that the measuring device can have variable speed with respect to the light beam. This violates the light postulate and the relativity principle. 

Adherence to the relativity principle for the measurement of $c$ from Maxwell's equations (light postulate) leads to time dilation and length contraction. Adherence to the relativity principle for the measurement of $h$ from Planck’s radiation law (Planck postulate) leads to `average-only' projection for spin-$\frac{1}{2}$ particles. That is, while an SG measurement cannot produce an outcome of $\cos{\left(\theta \right)}$ at $\hat{b}$ per NPRF + $h$, our $\pm 1$ ($\pm \frac{\hbar}{2}$) results can \textit{average} to the expected $\cos{\left(\theta \right)}$. Indeed, this is exactly what QM gives us. Here is how \cite{darrigol2022} justifies `average-only' projection:
\begin{quote}
    This is so because by a correspondence argument we expect the total angular momentum (or magnetic moment) of a large number of identically prepared, non-interacting spin-particles to behave as the angular momentum of a macroscopic object under measurement.
\end{quote}
This is also justified by the: 
\begin{quote}
    closeness requirement: the dynamics of a single elementary system can be generated by the invariant interaction between the system and a ``macroscopic transformation device'' that is itself described within the theory in the macroscopic (classical) limit.
\end{quote}
of \cite{dakic2013,dakicgroup} and the ``homogeneity of statistical ensembles'' per \cite{comte1996}. That is, the measuring devices used to measure quantum systems are themselves made from quantum systems. For example, \cite{bruknergroup} notes that the classical magnetic field of an SG magnet is used to measure the spin of spin-$\frac{1}{2}$ particles and that classical magnetic field ``can be seen as a limit of a large coherent state, where a large number of spin-$\frac{1}{2}$ particles are all prepared in the same quantum state.''

According to QM, to get the probability we have to square the probability amplitude. Specifically, we project $|\psi\rangle$ onto $|b+\rangle$ and square to obtain the probability that an SG spin measurement of $|\psi\rangle = |z+\rangle$ along $\hat{b}$ will produce $+1$. Likewise, we project $|\psi\rangle$ onto $|b-\rangle$ and square to obtain the probability that an SG spin measurement of $|\psi\rangle$ along $\hat{b}$ will produce $-1$. 

From Figures \ref{Qubit} and \ref{HilbertSpace} we see that the angle between $\hat{z}$ (for $|\psi\rangle = |z+\rangle$) and $\hat{b}$ (for $|b+\rangle$) is $\theta$ while the angle between $|z+\rangle$ and $|b+\rangle$ in Hilbert space is $\frac{\theta}{2}$. So, per QM we have $P(b+|\theta) = \cos^2{\left(\frac{\theta}{2}\right)}$ and $P(b-|\theta) = \sin^2{\left(\frac{\theta}{2}\right)}$ giving an average (expectation value) of 
\begin{center}
    $(+1)\cos^2{\left(\frac{\theta}{2}\right)} + (-1)\sin^2{\left(\frac{\theta}{2}\right)} = \cos{\left(\theta \right)}$
\end{center}
Conversely, one could use the Planck postulate and the closeness requirement to demand
\begin{center}
    $(+1)P(b+|\theta) + (-1)P(b-|\theta) = \cos{\left(\theta \right)}$.
\end{center}
With that equation and normalization
\begin{center}
    $P(b+|\theta) + P(b-|\theta) = 1$
\end{center}
we can then \textit{derive} the quantum-mechanical probabilities for our qubit.

\begin{figure}
\begin{center}
\includegraphics [height = 60mm]{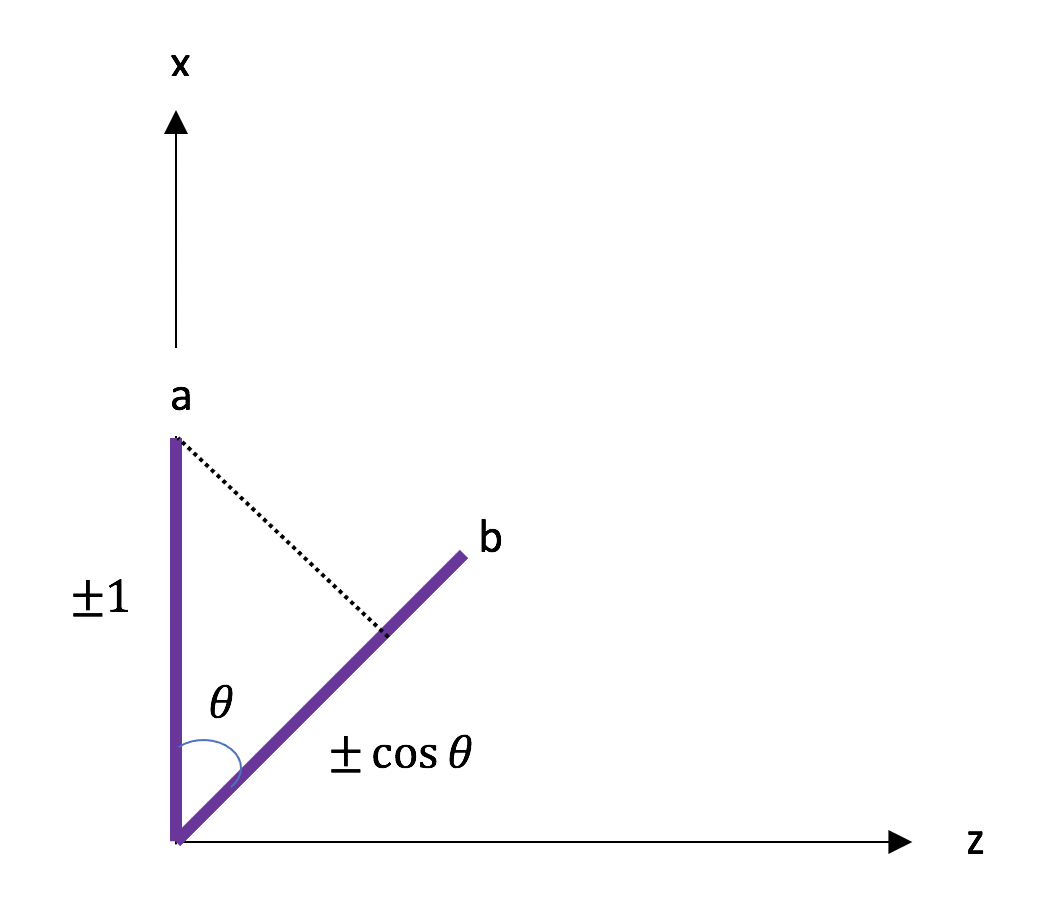}  \caption{Per Alice, Bob should be measuring $\pm \cos{\left(\theta \right)}$ when she measures $\pm1$, respectively.} \label{Alice-View}
\end{center}
\end{figure}

\begin{figure}
\begin{center}
\includegraphics [width = \textwidth]{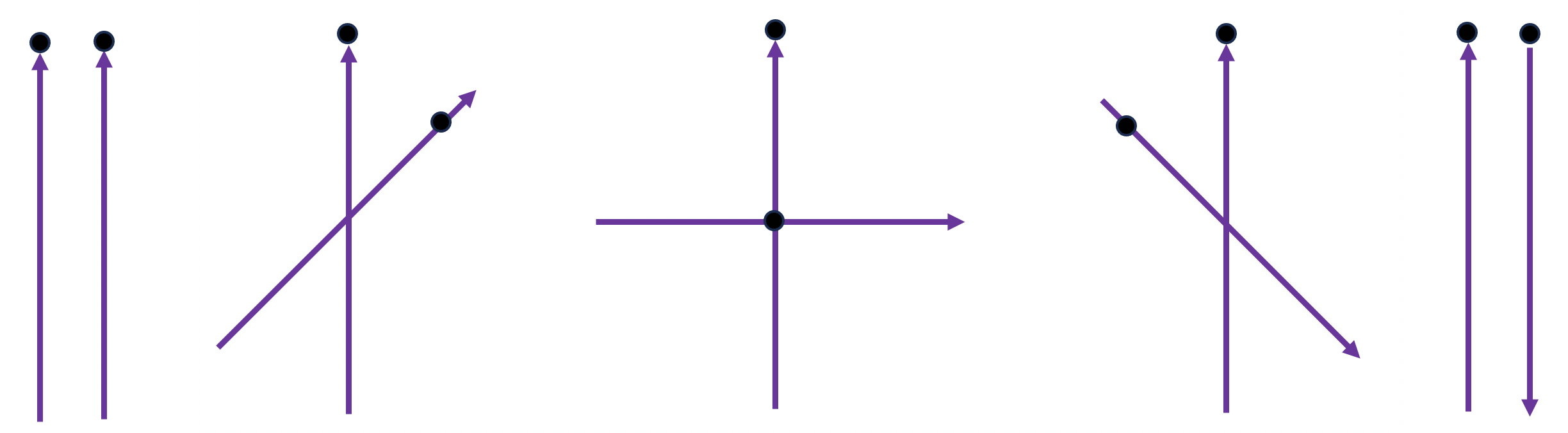}  \caption{\textbf{Average View for the Triplet State}. Reading from left to right, as Bob rotates his SG magnets (rotating purple arrow) relative to Alice's SG magnets (purple arrow always vertically oriented) for her $+1$ outcome (black dot at tip of her arrow), the average value of his outcome (black dot along his arrow) varies from $+1$ (totally up, arrow tip) to $0$ to $-1$ (totally down, arrow bottom). This obtains per conservation of spin angular momentum on average in accord with the relativity principle. Bob can say exactly the same about Alice's outcomes as she rotates her SG magnets relative to his SG magnets for his $+1$ outcome.} \label{AvgViewTriplet}
\end{center}
\end{figure}

Now suppose you have a pair of qubits in the symmetry plane of a Bell spin triplet state and Alice obtains $+1$ at $\hat{a}$ and Bob measures his particle at $\hat{b} \ne \hat{a}$ ($\theta \ne 0$). We have the exact same situation here between $\hat{a}$ and $\hat{b}$ for two particles that we had between $\hat{z}$ and $\hat{b}$ for one particle. Using the same reasoning, Alice says Bob's measurement outcome should be $\cos{\left(\theta \right)}$, since obviously he would have also gotten $+1$ for his particle if he had measured at $\hat{b} = \hat{a}$, as required to conserve spin angular momentum per the spin triplet state (Figure \ref{Alice-View}). The problem is, again, that would mean Alice alone measures $h$ while Bob measures some fraction of $h$, which means Alice occupies a preferred reference frame. Since Bob must also always measure $h$ per the relativity principle, Bob's $\pm 1$ outcomes can only \textit{average} to $\cos{\left(\theta \right)}$ at best (Figure \ref{AvgViewTriplet}). That means from Alice's perspective, Bob's measurement outcomes only satisfy conservation of spin angular momentum \textit{on average} when Bob is measuring the spin of his particle in a different inertial reference frame. 

We can write this `average-only' conservation for Alice's $+1$ outcomes as 
\begin{center}
$2P(++)(+1) + 2P(+-)(-1) = \cos{\left(\theta \right)}$.
\end{center}

\clearpage

\noindent Likewise, for Alice's $-1$ outcomes `average-only' conservation is written
\begin{center}
$2P(-+)(+1) + 2P(--)(-1) = -\cos{\left(\theta \right)}$.
\end{center}
This `average-only' conservation plus normalization per the relativity principle:
\begin{align*}
P(++) + P(+-) & = \frac 12 \\
P(-+) + P(--) & = \frac 12
\end{align*}
gives the joint probabilities $P(++) = P(--) = \frac{1}{2}\cos^2{\left(\frac{\theta}{2}\right)}$ and $P(+-) = P(-+) = \frac{1}{2}\sin^2{\left(\frac{\theta}{2}\right)}$ for the Bell spin triplet state in its symmetry plane in accord with QM. 

\begin{figure}
\begin{center}
\includegraphics [width = \textwidth]{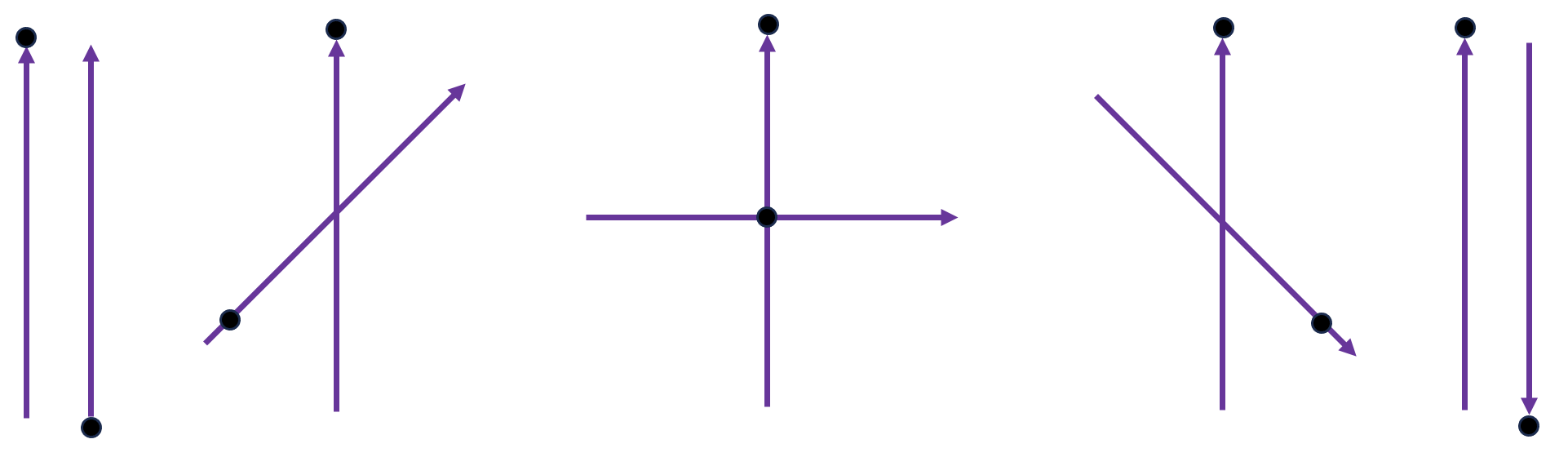}  
\caption{\textbf{Average View for the Singlet State}\index{Bell states}. Compare with Figure \ref{AvgViewTriplet}.} \label{AvgViewSinglet}
\end{center}
\end{figure}

All of these arguments can be applied equally to the singlet state. There the probabilities to be derived are reversed: $P(+-) = P(-+) = \frac{1}{2}\cos^2{\left(\frac{\theta}{2}\right)}$ and $P(++) = P(--) = \frac{1}{2}\sin^2{\left(\frac{\theta}{2}\right)}$. Since Bob obtains $-1$ when Alice obtains $+1$ for $\hat{b} = \hat{a}$ and vice-versa for the singlet state, `average-only' conservation for Alice's $+1$ outcomes becomes 
\begin{center}
$2P(++)(+1) + 2P(+-)(-1) = -\cos{\theta}$
\end{center}
while for Alice's $-1$ outcomes `average-only' conservation becomes
\begin{center}
$2P(-+)(+1) + 2P(--)(-1) = \cos{\left(\theta \right)}$.
\end{center}
Normalization per NPRF is the same, so normalization and these properly revised `average-only' conservation equations do in fact give the singlet state joint probabilities. Consequently Figure \ref{AvgViewTriplet} now looks like Figure \ref{AvgViewSinglet}.

\begin{figure}
\begin{center}
\includegraphics [height = 60mm]{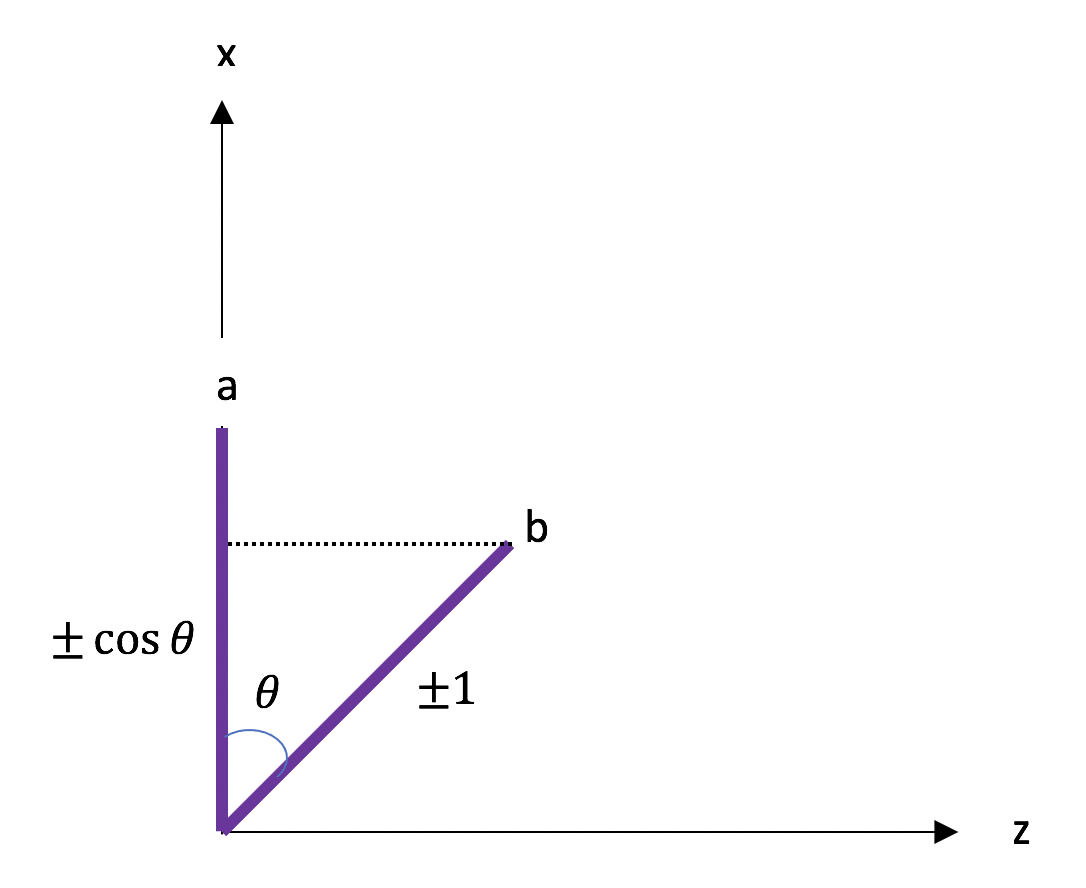}  \caption{Per Bob, Alice should be measuring $\pm \cos{\left(\theta \right)}$ when he measures $\pm1$, respectively.} \label{Bob-View}
\end{center}
\end{figure}

\begin{figure}
\begin{center}
\includegraphics [width = \textwidth]{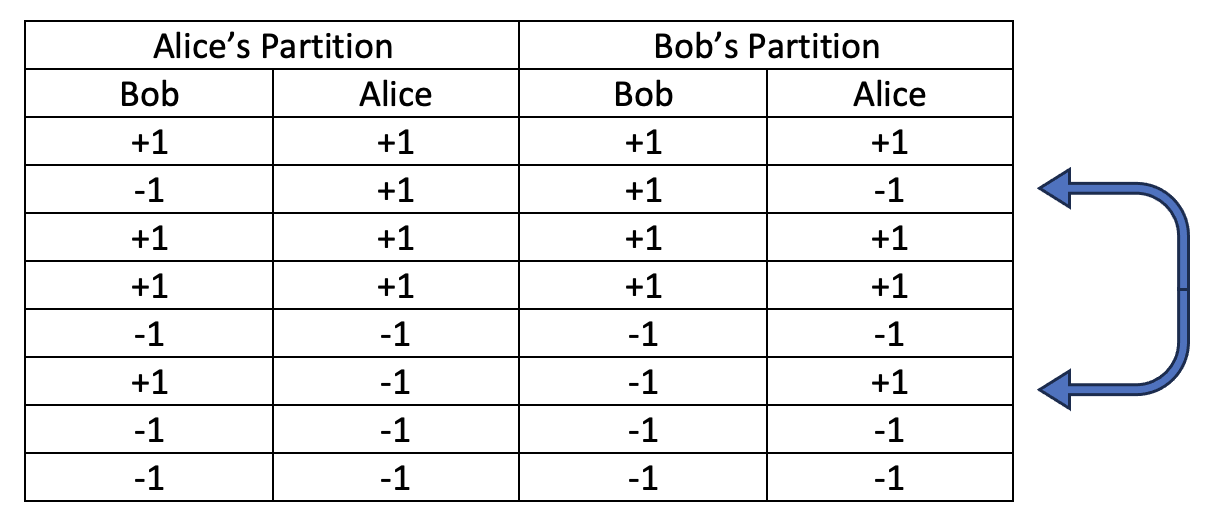}  \caption{Example collection of eight data pairs when Alice and Bob's measurement settings in the symmetry plane for a Bell spin triplet state differ by $60^{\circ}$. Alice partitions the data according to her $\pm 1$ results to show that Bob's measurement outcomes only \textit{average} the required $\pm\frac{1}{2}$ for conservation of spin angular momentum. But, when Bob partitions the data according to his $\pm 1$ results (by switching rows 2 and 6 shown with blue arrows) he can show it is \textit{Alice's} measurement outcomes that only average the required $\pm\frac{1}{2}$ for conservation of spin angular momentum.} \label{QGdata}
\end{center}
\end{figure}

Of course the data are symmetrical, so Bob can partition the data according to \textit{his} $+1$ and $-1$ results and argue equally that it is Alice who must average \textit{her} data in accord with the conservation of spin angular momentum (Figures \ref{AvgViewTriplet}, \ref{Bob-View} and \ref{QGdata}). This is exactly analogous to SR's relativity of simultaneity between reference frames in uniform relative motion. There Alice partitions events in spacetime according to her surfaces of simultaneity and says that Bob's meter sticks are short and his clocks run slow, while Bob partitions events in spacetime according to his surfaces of simultaneity and says that it is Alice's meter sticks that are short and her clocks that run slow.

\clearpage

For vertically polarized light passed through a polarizer angled at $45^{\circ}$ with respect to the vertical, classical electromagnetism says half the light will pass. So, when we send vertically polarized photons to the polarizer, half will pass and half will not. If `half photons' passed through the polarizer in accord with classical physics, then their energy on the other side would be given by $E = \frac{h}{2}f$ instead of $E = hf$, meaning the value of Planck's constant was effectively cut in half \cite{PolarizersMIT}. 

So, the classical `expectation' of fractional amounts of quanta holds on average per Information Invariance \& Continuity. That is, we would have `average-only' transmission of energy for spin-1 photons instead of `average-only' projection of spin angular momentum for spin-$\frac{1}{2}$ particles, both of which give `average-only' conservation between different inertial reference frames related by spatial rotations \cite{stuckey2019,stuckey2020}.

In the double-slit experiment, the Planck postulate demands we obtain quanta of momentum $p = \frac{h}{\lambda}$ in our detector at any location along the screen. If that momentum was instead distributed in a continuum of wave intensity along the screen as in classical optics, then our detector would be getting a fraction of $h$ in $p = \frac{h}{\lambda}$ at any given location along the screen. However, in the quantum measurement context for momentum $p$, an interference pattern allowing us to compute $\lambda$ for use in $p = \frac{h}{\lambda}$ is required. Therefore, the interference pattern per classical wave mechanics must obtain on average from the distribution of discrete quantum detection events along the detector screen. In the double-slit experiment, the mystery of the qubit (superposition/noncommutativity/complementarity) resides in this wave-particle duality \cite{NPRF2024}.  

In all three examples, NPRF + $h$ demands that a classically continuous quantity (angular momentum, energy, or momentum) be quantized, so that the classically continuous prediction obtains on average over the distribution of quantum events. Everything said here is consistent with the standard textbook presentation of the quantum-classical relationship. We're simply pointing out that you can think of QM as following from the Planck postulate as justified by the relativity principle, such that this quantum-classical relationship obtains. 

Essentially, NPRF + $c$ constrains the spacetime configuration of worldtubes for bodily objects (with their unlimited simultaneous information) according to the invariant and large-but-finite speed $c$ of information transfer while NPRF + $h$ dictates the distribution of quanta (with their limited simultaneous information) among those bodily objects according to the invariant and small-but-nonzero action $h$. Thus, NPRF + $c$ and NPRF + $h$ can be understood as adynamical global constraints over the distribution of quantum events in the context of bodily objects in spacetime in accord with ``all-at-once'' explanation as explained in Section \ref{SectionIntro}.

\clearpage


\section{Conclusion}\label{SectionConcl}

\cite{BerghoferIQOQI2023} echoed Rovelli's challenge in a talk at IQOQI:
\begin{quote}
    One crucial motivation for QRP is that Einstein did something similar for special relativity. The mathematics underlying special relativity was discovered before Einstein, but only when Einstein succeeded in deriving the mathematics from meaningful principles -- the light postulate and the principle of relativity -- did special relativity emerge as a well-understood and broadly accepted scientific theory. 
\end{quote}
Herein, we argued that QRP has rendered QM a principle theory in total analogy with SR. Specifically, Information Invariance \& Continuity provides us with an empirically discovered fact whence the Hilbert space formalism of QM in complete analogy with the light postulate providing an empirically discovered fact whence the Lorentz transformations of SR. However, QRP failed to provide an understanding of QM as robust as that of SR because structural explanation via Hilbert space does not provide ``clear empirical understanding'' or unification on par with that of the Lorentz transformations. 

Therefore, we proposed thinking of SR as a principle explanation rather than a structural explanation when making comparisons with QM. To make this possible, we showed how Information Invariance \& Continuity with spatialized measurement entails that everyone measure the same value for Planck's constant $h$ (from Planck’s radiation law), regardless of their relative spatial orientations or locations (Planck postulate) in exact analogy with the fact that everyone measures the same value for the speed of light $c$ (from Maxwell's equations), regardless of their relative velocities (light postulate). Since both the Planck and light postulates can be justified by the relativity principle, both QM and SR are understood as principle explanations with ``clear empirical understanding'' and unification much broader than SR alone (Figure \ref{UnificationBig}). 

\begin{figure}
\begin{center}
\includegraphics [width = \textwidth]{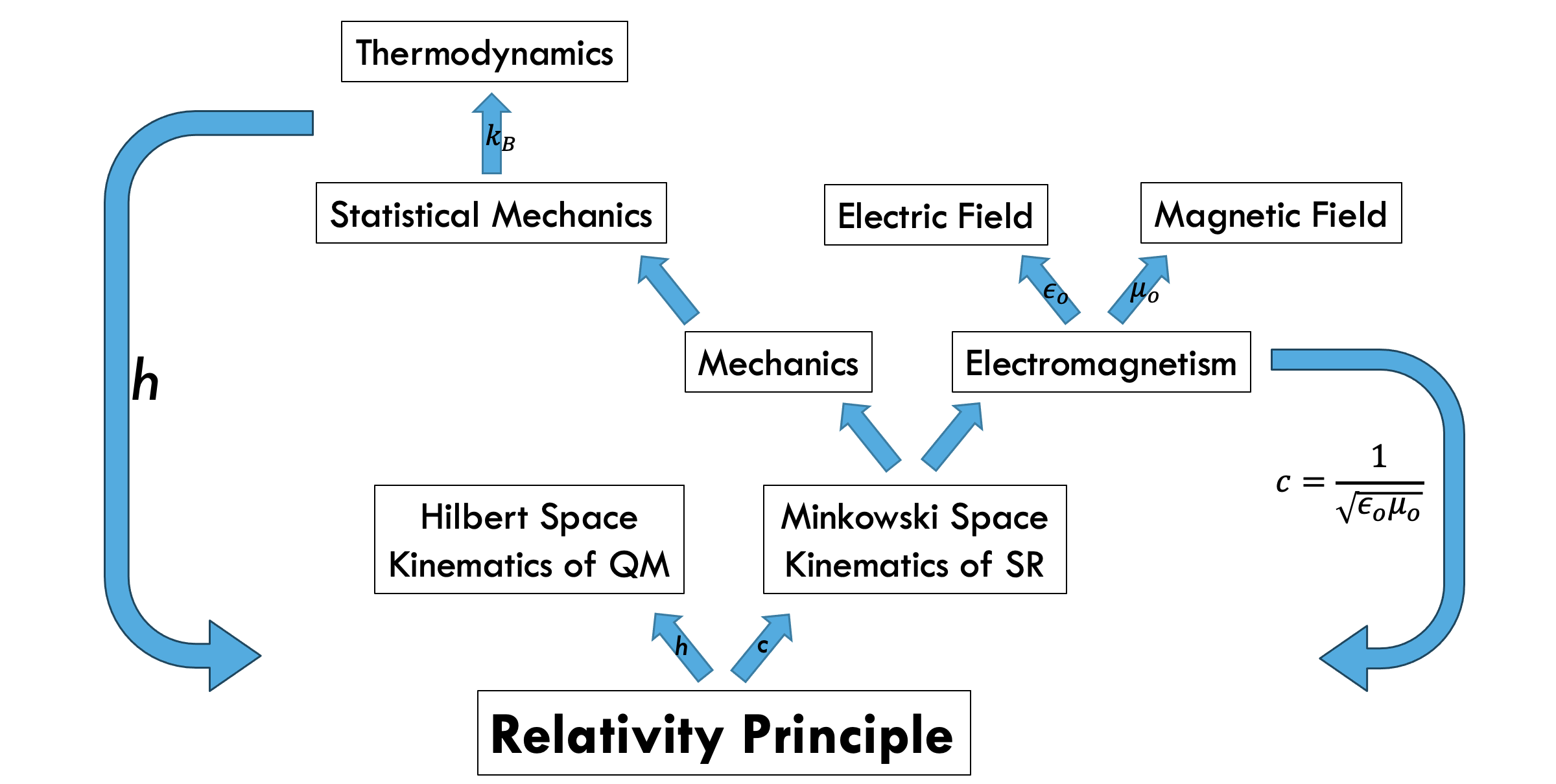}  
\caption{\textbf{Kinematics constrain dynamics, which bear on kinematics}.} \label{DynamicsOnKinematics}
\end{center}
\end{figure}

We should point out that while the kinematics constrain the dynamics, the dynamics obviously bear on the kinematics. As \cite{PlanckRadiationLaw} and \cite{PlanckRadiationYouTube} explain, Planck first derived his constant $h$ using thermodynamics and electromagnetism to fit Wien's law to existing experimental data (a year before he found his radiation law, eventually adding statistical mechanics to his approach) as a necessary precursor to QM. And of course, Maxwell derived $c$ from his theory of electromagnetism as a necessary precursor to SR. So, Figure \ref{UnificationBig} is more accurately given by Figure \ref{DynamicsOnKinematics}. Hilbert space and M4 are the kinematics for low-energy situations and they are changed dramatically when one is dealing with high-energy dynamics. 

As mentioned in Section \ref{SectionIntro}, Schr\"odinger's equation of QM is only the low-energy approximation to the Lorentz-invariant Klein-Gordon equation of quantum field theory with a kinematics of Fock space, a direct sum over multi-particle Hilbert spaces. And in SR, M4 spacetime is globally flat and independent of the stress-energy tensor. But, we know that M4 is only a low mass-energy approximation to the curved spacetime of general relativity, which is determined by the stress-energy tensor per Einstein's equations and is only locally flat. So in general, kinematics constrain dynamics, which in turn bear prominently on the kinematics (beyond the simple situation shown in Figure \ref{DynamicsOnKinematics} for low-energy situations). 

We also explained how the Planck postulate is the reason for the mysterious superposition/noncommutativity/complementarity whence `average-only' projection for spin-$\frac{1}{2}$ particles, `average-only' transmission for photon polarization, and `average-only' intensity for the double-slit experiment. Accordingly, the most fundamental explanans for Bell state entanglement is `average-only' conservation resulting from the relativity principle. Likewise, rather than using the geometry of M4 as the fundamental explanans for the mysteries of length contraction and time dilation per structural explanation, the most fundamental explanans for length contraction and time dilation per principle explanation is the relativity of simultaneity resulting from the relativity principle. Consequently, \cite{koberinski2018} write:
\begin{quote}
    We suggest that (continuous) reversibility may be the postulate which comes closest to being a candidate for a glimpse on the genuinely physical kernel of ``quantum reality.'' Even though Fuchs may want to set a higher threshold for a ``glimpse of quantum reality,'' this postulate is quite surprising from the point of view of classical physics: when we have a discrete system that can be in a finite number of perfectly distinguishable alternatives, then one would classically expect that reversible evolution must be discrete too. For example, a single bit can only ever be flipped, which is a discrete indivisible operation. Not so in quantum theory: the state $|0\rangle$ of a qubit can be continuously-reversibly ``moved over'' to the state $|1\rangle$. For people without knowledge of quantum theory (but of classical information theory), this may appear as surprising or ``paradoxical'' as Einstein's light postulate sounds to people without knowledge of relativity.
\end{quote}

The analogy between QM and SR is historical as well. Physicists (including Einstein) first tried to find a causal mechanism for length contraction a la the luminiferous aether before Einstein ``despairingly'' gave up his ``constructive efforts'' and turned to a principle approach (relativity principle) resulting in SR as we know it today. Likewise, physicists first tried to find a causal mechanism for Bell state entanglement before (and while) Rovelli, Fuchs, Hardy, Zeilinger, Brukner, M\"uller, and others turned to a principle approach based on information-theoretic principles. Viewing QRP as a ``device-independent approach'' where physical systems need not have counterparts in the operational formalism, \cite{Grinbaum2017} writes:
\begin{quote}
    If, despite Einstein's wish, no constructive theory has materialized as a replacement of special relativity, it is not impossible to imagine that our intuitive desire to `fill the box' with physical systems for the purposes of better explaining physics is as illusory. The device-independent approach might stay as a legitimate way of doing physics, without any need to `fill the box,' much in the same sense as principle-based special relativity has not been surpassed by any constructive theory.
\end{quote} 
QRP has succeeded in rendering QM a principle theory, but not a principle explanation a la SR, so it has not found consensus support like SR.

Another reason QRP's principle version of QM has not found consensus support in foundations is due to a pervasive constructive bias in foundations. For example, confronted with the double-slit experiment (what \cite[ch. 37]{FeynmanDoubleSlit} called ``the \textit{only} mystery'' of QM) the constructive thinker wants to know exactly what is passing through the slits. We end up with particle-like outcomes at the detector that are distributed in a wave-like interference pattern, so are particles passing through the slits or are waves? If it's particles, then they must pass through one slit or the other, so how do they end up creating an interference pattern? Perhaps you have waves guiding particles? If it's waves, then they impinge along the entirety of the detector screen, so how do they end up creating particle events? Perhaps it's spontaneous collapse? Or perhaps the waves don't collapse, but the universe splits into infinitely many worlds each with a point-like outcome?

Here is a typical constructive attitude about the double-slit experiment from Maudlin \cite{maudlin2023}:
\begin{quote}
You would like a clear physical account of what's really going on. ... The first thing you want to ask somebody when they're trying to explain the physics underlying this behavior is, ``Do you have the first picture where in addition to the particle you have this wave-like thing (like Bohmian mechanics) or the second picture where you only have this wave-like thing and it collapses (spontaneous collapse models) or the third picture where you only have this wave-like thing and it doesn't collapse (like Many-Worlds).'' Those are your three options.
\end{quote}
It may be true that those are your only three options if you are looking for a constructive solution to the mystery of the double-slit experiment. But, as we explained above, there is an entirely different way to solve the mystery of the double-slit experiment. That is, the most fundamental explanans for wave-particle duality in the double-slit experiment is `average-only' intensity due to the relativity principle demanding the observer-independence of $h$ between inertial reference frames related by spatial translations. And, it's a matter of opinion as to whether or not such principle explanation is necessarily inferior to the constructive explanation of the interpretation program. For example, \cite{koberinski2018} retort:
\begin{quote}
    None of Bohmian mechanics, Everettian quantum theory, or collapse theories fill the explanatory role of a principle theory. For example, Everettian quantum theory does not start with a broad general framework of `theories of many worlds,' put simple principles on top of that, and prove that quantum theory is the unique theory of many worlds that satisfies these principles.
\end{quote}

\clearpage

\begin{figure}
\begin{center}
\includegraphics [width = \textwidth]{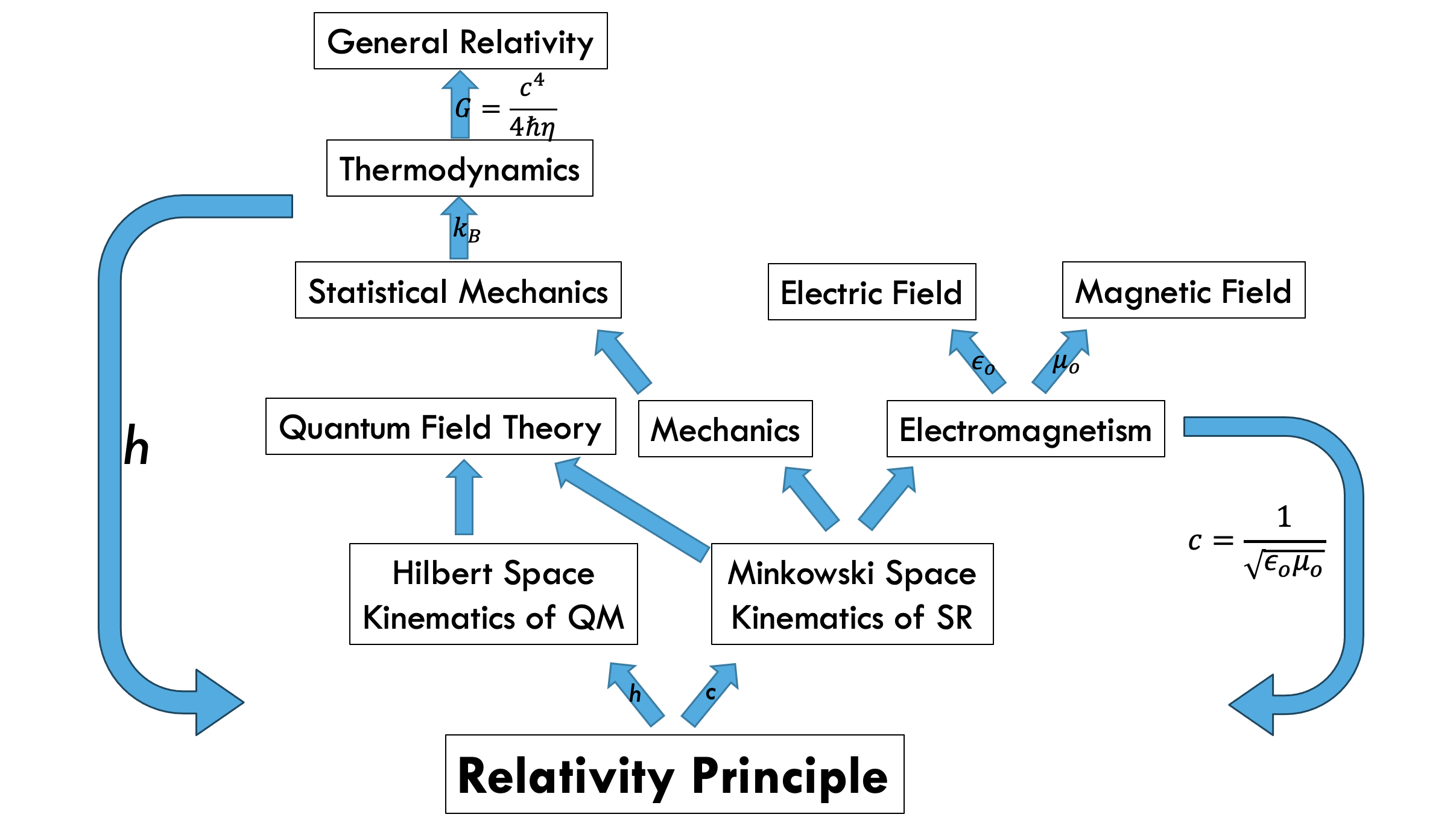}  
\caption{\textbf{All Physics}.} \label{UnificationBiggest}
\end{center}
\end{figure}

Additionally, as we have shown herein, the completion of QRP via principle explanation (based on a spatial interpretation of QRP) provides enormous unification (Figure \ref{DynamicsOnKinematics}) with ``clear empirical understanding.'' If you take into account the \cite{Jacobson1995} derivation of Einstein's equations from thermodynamic principles and the equivalence principle (itself a form of NPRF), combined with the fact that quantum field theory resulted from the marriage of QM and SR, then Figure \ref{DynamicsOnKinematics} becomes Figure \ref{UnificationBiggest}. [Aside: According to Jacobson's derivation, Newton's gravitational constant $G = \frac{c^4}{4\hbar\eta}$, where $\eta$ is the constant of proportionality between entropy and causal horizon area.] And, this view solves the mystery of entanglement without violating locality, statistical independence, intersubjective agreement, or the uniqueness of experimental outcomes, which is provably impossible in constructive fashion per Bell's theorem and the experimental violation of Bell inequalities. Thus, QM and SR are unified via adynamical global constraints such that both are well-understood principle theories despite the fact that neither has a consensus constructive counterpart. What more might the Nobel laureate in physics require in order to say they now understand QM?


\clearpage

\bibliographystyle{apalike}
\bibliography{biblio} 
\end{document}